\documentclass[a4paper,11pt]{article}

\usepackage{natbib}
\usepackage{eqnarray,amsmath,amsthm}
\usepackage{amssymb}
\usepackage{graphicx}
\usepackage{epstopdf}
\usepackage{authblk} 
\usepackage{mathtools}
\usepackage{mathabx}
\usepackage{xcolor,colortbl}
\usepackage{algorithm, algorithmicx}
\usepackage{algpseudocode}
\usepackage{changepage}
\usepackage{subfigure}
\usepackage[margin=3cm]{geometry}
\usepackage{float}
\usepackage{nicefrac}
\usepackage{hyperref}
\usepackage{microtype} 
\usepackage{enumitem}

\newcommand{\keywords}[1]{\textbf{\textit{Keywords---}} #1}
\newcommand{\norm}[1]{\left\lVert#1\right\rVert}
\newcommand{\vect}[1]{\boldsymbol{#1}}
\newcommand{\params}{\boldsymbol{\theta}}
\newcommand{\ceil}[1]{\left\lceil #1\right\rceil}
\newcommand{\Nt}{N_{\theta}}
\newcommand{\Nx}{N_{x}}
\newcommand{\ESS}{\textrm{ESS}}
\newcommand{\Eff}[1]{\textrm{Eff}_{\textrm{#1}}}
\newcommand{\RelEff}[1]{\textrm{RelEff}_{\textrm{#1}}}

\makeatletter
\algnewcommand{\LineComment}[1]{{\Statex \hskip\ALG@thistlm \footnotesize\textcolor{blue}{/* #1 */}}}
\algnewcommand{\LineCommentIndent}[1]{{\Statex \hskip\ALG@tlm \footnotesize\textcolor{blue}{/* #1 */}}}
\makeatother

\title{\bf Adaptively switching between a particle marginal Metropolis-Hastings and a particle Gibbs kernel in SMC$^2$}
\author[$1$,$3$,$4$]{Imke Botha} 
\author[$2$,$3$,$5$]{Robert Kohn}
\author[$1$,$3$,$4$]{Leah South}
\author[$1$,$3$,$4$]{Christopher Drovandi}

\affil[$1$]{School of Mathematical Sciences, Queensland University of Technology (QUT)}
\affil[$2$]{School of Economics, University of New South Wales}
\affil[$3$]{Australian Research Council Centre of Excellence for Mathematical \& Statistical	Frontiers}
\affil[$4$]{QUT Centre for Data Science}
\affil[$5$]{DARE: ARC training centre in data analytics for reources and environments}

\begin{document}
	\setlength{\parindent}{0pc}
	\setlength{\parskip}{1ex}
	\maketitle
	
	\begin{abstract}
		Sequential Monte Carlo squared \citep[SMC$^2$; ][]{Chopin2012} methods can be used to sample from the exact posterior distribution of intractable likelihood state space models. These methods are the SMC analogue to particle Markov chain Monte Carlo \citep[MCMC; ][]{Andrieu2010} and rely on particle MCMC kernels to mutate the particles at each iteration. Two options for the particle MCMC kernels are particle marginal Metropolis-Hastings (PMMH) and particle Gibbs (PG). We introduce a method to adaptively select the particle MCMC kernel at each iteration of SMC$^2$, with a particular focus on switching between a PMMH and PG kernel. The resulting method can significantly improve the efficiency of SMC$^2$ compared to using a fixed particle MCMC kernel throughout the algorithm. Code for our methods is available at \url{https://github.com/imkebotha/kernel_switching_smc2}.
	\end{abstract}
	
	
	\section{Introduction}
	
	Sequential Monte Carlo squared \citep[SMC$^2$;][]{Chopin2012,Duan2014} can be used to sample from the posterior distribution of state space models (SSMs) where the likelihood function of the data given the model parameters is intractable. It is the SMC analogue to particle Markov chain Monte Carlo \citep[MCMC;][]{Andrieu2009,Andrieu2010}, and retains the following advantages of SMC over MCMC: it is inherently parallelisable, it is better able to detect multimodality, it estimates the marginal likelihood as a by-product, and it allows for tuning parameters to be adapted automatically. SMC$^2$ has been applied in a number of areas including macroeconomics and ﬁnance \citep{Fulop2022}, epidemiology \citep{Golightly2017,Rimella2022} and paleoclimatology \citep{Carson2017}.
	
	For static models, SMC propagates a set of parameter particles through a sequence of distributions linking the prior to the posterior \citep{Chopin2002,DelMoral2006}. Samples are initially drawn from the prior, then reweighted, resampled and mutated for each subsequent distribution in the sequence. In SMC$^2$, a particle MCMC kernel, typically particle marginal Metropolis Hastings (PMMH) or particle Gibbs (PG) \citep{Andrieu2010}, is used to mutate the particles. PMMH uses a particle filter estimator of the likelihood within a Metropolis-Hastings algorithm, and PG uses a conditional particle filter to unbiasedly draw a latent state trajectory from its full conditional distribution. The model parameters are then updated conditional on this trajectory.  
	
	Depending on the particular model and application, PMMH or PG may be preferred. PMMH can be used when the transition density of the SSM is analytically intractable, and may be preferred over PG if the latent states and the parameters are highly correlated or if there is path degeneracy in the particle filter \citep{Chopin2015,Svensson2015}, i.e.\ when most of the particles share a common ancestor. Path degeneracy is more common with longer time series, and it can significantly restrict movement in the sampled latent state trajectories, particularly at the initial time points. PG generally requires that the transition density is tractable analytically, but it allows for more flexibility in the update of the model parameters. For example, it is straightforward to update the model parameters jointly, in blocks or singly and more sophisticated algorithms can be used, such as the Metropolis-adjusted Langevin algorithm \citep[MALA;][]{Girolami2011} or Hamiltonian Monte Carlo \citep[HMC;][]{Betancourt2017}. Furthermore, PG variants that mitigate issues arising from path degeneracy have been developed \citep{Whiteley2010,Lindsten2012,Lindsten2014}.
	
	The efficiency of particle MCMC methods, and consequently SMC$^2$, greatly depends on the number of state particles $N_x$ used within the particle filter. For both PMMH and PG, $N_x$ must scale linearly with the time series \citep{Andrieu2010}. A general recommendation for PMMH is to select $N_x$ such that the variance of the log of the likelihood estimator is between 1 and 3 at, for example, the posterior mean \citep{Pitt2012,Doucet2015,Sherlock2015}. Methods to adaptively update $N_x$ when using PMMH within SMC$^2$ are available \citep{Chopin2012,Chopin2015a,Duan2014,Botha2023}, but currently there is no straightforward method to select $N_x$ for PG. In the context of MCMC, however, it has been observed that PG typically requires much fewer particles than PMMH to achieve good performance \citep[see e.g.\ Section 16.5.2 of][]{Chopin2020}. 
	
	Choosing whether to use PMMH or PG in SMC$^2$ is further complicated by the sequence of distributions being traversed. At each iteration, the density targeted by SMC$^2$ changes, and only at the final iteration is the target the posterior distribution. As a result, the optimal choice of particle MCMC kernel, in terms of computational and statistical efficiency, may change across iterations. Additionally, while PG may be less expensive than PMMH due to a lower value of $N_x$, it may require more intermediate distributions\slash targets due to the dependence on the latent state trajectory in the reweighting step. 
	
	Our paper introduces an automatic method to choose a particle MCMC kernel at each iteration in SMC$^2$. We motivate and test our approach using a PMMH and a PG kernel; however, our proposed approach can straightforwardly be extended to any set of candidate kernels, for example, three different PMMH kernels or a PMMH kernel and two PG kernels. In some settings, our framework allows the latent states to be integrated out during the reweighting step, while still using PG to mutate the particles. Notably, it also allows for a different number of state particles to be used for PMMH and PG. 
	
	The rest of the paper is organised as follows. Section \ref{sec:background} gives the background on SSMs, SMC and SMC$^2$. Sections \ref{sec:switching_mechanics} and \ref{sec:adaptation} outline our proposed approach for switching between PMMH and PG kernels in SMC$^2$. Section \ref{sec:examples} describes the implementation of our numerical experiments, including the particular particle MCMC kernels that we use, and shows the performance of our approach on a Brownian motion model, a flexible-allee logistic population model, a stochastic volatility in mean model and an AR(1) model. Section \ref{sec:discussion} concludes with a discussion and possible future work.
	
	\section{Background} \label{sec:background}	
	We use the colon notation for collections of random variables, i.e.\ $z_{i:j} := \{z_i, z_{i+1}, \ldots, z_j\}$, $z^{k:m} := \{z^k, z^{k+1}, \ldots, z^m\}$ and $z_{i:j}^{k:m} := \{z_i^{k:m}, z_{i+1}^{k:m}, \ldots, z_j^{k:m}\}$. Similarly, for the exclusion of the $l$th element we use $z^{k:m, -l} := \{z^{k}, z^{k+1}, \ldots, z^{l-1}, z^{l+1} \ldots, z^{m}\}$. Boldface is used to denote a collection in the absence of an identifying subscript or superscript, e.g\ $\vect{w}$, $\vect{v}_i$ and $z_{i:j}$ are all collections. 
	
	\subsection{State Space Models} \label{sec:ssms}	
	We consider a state space model with parameters $\params$, latent states $X_{1:T} = x_{1:T}$ and observations $Y_{1:T} = y_{1:T}$. The posterior distribution of the unknown parameters $\params$ is 
	\begin{align}
		p(\params\mid y_{1:T})	= \frac{p(\params)p(y_{1:T}\mid\params)}{p(y_{1:T})} = \frac{p(\params)}{p(y_{1:T})} \int{
				p(x_{1:T}, y_{1:T}\mid \params)
			}dx_{1:T},
	\label{eqn:SSM_posterior}
	\end{align}
	and the joint density of the latent states and observations is
	\begin{align}
		p(x_{1:T}, y_{1:T}\mid \params) = \mu(x_1\mid\params)\prod_{t=2}^T{f(x_t \mid x_{t-1}, \params)}\prod_{t=1}^T{g(y_t\mid x_t, \params)},
		\label{eqn:posterior}
	\end{align}
	where $g(y_t\mid x_t, \vect\theta)$ is the observation density, $f(x_t\mid x_{t-1}, \vect\theta)$ is the transition density, and $\mu(x_1\mid\params)$ is the density of the initial state.
	
	The likelihood $p(y_{1:T}\mid\params)$ in \eqref{eqn:SSM_posterior} is intractable if the integral is analytically intractable or too expensive to compute, or if the transition density $f(x_t \mid x_{t-1}, \params)$ is unknown in closed form. For a fixed $\params$, however, a particle filter can be used to obtain an unbiased estimate of $p(y_{1:T}\mid\params)$ \citep{DelMoral2006}. Likewise, a conditional particle filter can be used to unbiasedly draw a latent state trajectory from its full conditional distribution \citep{Andrieu2010}. The next section describes SMC methods, including their application to dynamic models, i.e.\ particle filters. 
	
	\subsection{Sequential Monte Carlo} \label{sec:smc}	
	SMC methods sample through the sequence of distributions given by
	\begin{align}
		\pi_d(z_d) = \frac{\gamma_d(z_d)}{V_d} \ \propto \ \gamma_d(z_d), \quad d = 0, \ldots, D,
		\label{eqn:smc_targets}
	\end{align}
	where $V_d$ is the unknown normalising constant, $\pi_0(z_0)$ is the initial distribution, and $\pi_D(z_D)$ is the distribution of interest. Initially, $N_z$ evenly-weighted particles are drawn from the initial distribution, $z_0\sim \pi_0(z_0)$, giving $\{z_0^{n}, W_0^n=\nicefrac{1}{N_{z}}\}_{n=1}^{N_{z}}$. Then, for each subsequent distribution in the sequence, the weighted particles are mapped from $\pi_{d-1}(z_{d-1})$ to $\pi_{d}(z_{d})$, for $d = 1, \ldots, D$, through the following steps:
	\begin{enumerate}[label={(\arabic*)}]
		\item Resample the particles according to their weights. \label{enum:step1}
		\item Mutate the particles using a mutation kernel $K(z^n_{d-1},z^n_{d})$ for $n = 1, \ldots, N_z$. \label{enum:step2}
		\item Reweight the particles using the ratio of the unnormalised targets, the mutation kernel $K(z^n_{d-1},z^n_{d})$ and an artificial backward kernel $L(z^n_{d}, z^n_{d-1})$ \citep{DelMoral2006}, \label{enum:step3}
		\begin{align*}
			w_d^n =  W_{d-1}^n\cdot\frac{\gamma_{d}(z^n_{d})L(z^n_{d}, z^n_{d-1})}{\gamma_{d-1}(z^n_{d-1})K(z^n_{d-1}, z^n_{d})}, \quad W_d^n = \frac{w_d^n}{\sum_{i=1}^{N_{z}}w_d^i}, \quad n = 1, \ldots, N_z.
		\end{align*}
	\end{enumerate}
	Step \ref{enum:step1} removes low-weight particles and duplicates high-weight particles, and Step \ref{enum:step2} aims to counteract particle depletion caused by Step \ref{enum:step1}. If the weights are independent of the mutated particles, i.e.\ $w_d^n$ is independent of $z^n_{d}$, then the order of the steps become \ref{enum:step3} $\rightarrow$\ref{enum:step1}$\rightarrow$\ref{enum:step2} \citep{DelMoral2006}. At each iteration, the weighted sample $\{z_d^n, W_d^n\}_{n=1}^{N_{z}}$ approximates $\pi_d(z_d)$. Degeneracy of the particles, i.e.\ when most of the weight is given to a few particles, can be measured using an estimate of the effective sample size (ESS), $\widehat{\ESS}_d = \nicefrac{1}{\sum_{i=1}^{\Nt}{(W_d^i)^2}}$ \citep{Liu1998}. In many cases, it is more efficient to only resample and mutate the particles (Steps \ref{enum:step1} and \ref{enum:step2}) when the ESS falls below some threshold $\ESS_{\textrm{threshold}}$, for example, $\ESS_{\textrm{threshold}} = \nicefrac{\Nt}{2}$. This is commonly referred to as adaptive resampling. 
	
	An attractive feature of SMC is that an unbiased estimate of the normalizing constant $V_D$ can be calculated using the unnormalized weights \citep{DelMoral2006}
	\begin{align}
		\int{\gamma_{D}(z_D)} dz \approx \prod_{d=1}^{D}{\sum_{n=1}^{N_{z}}{w_d^{n}}}.
		\label{eqn:llest}
	\end{align}
	The next two sections describe how SMC can be applied to static and dynamic models. 

	\subsubsection{Static models} \label{sec:smc_static}	
	Consider a static model with posterior
	\begin{align*}
		p(\params\mid y_{1:T}) = \frac{p(\params)p(y_{1:T}\mid\params)}{p(y_{1:T})},
	\end{align*}
	where posterior samples of $\params$ are of interest. Then $z = \params$, $\pi_D(\params_D) = 	p(\params\mid y_{1:T})$ and often the initial distribution is the prior, i.e.\ $\pi_0(\params_0) = p(\params)$. What remains to specify is the sequence of distributions linking $p(\params)$ to $p(\params\mid y_{1:T})$, the mutation kernel and the backward kernel. 
	
	Two common choices for constructing the sequence of distributions are data annealing 
	\begin{align*}
		\pi_d(\params_d) \ \propto \ p(\params)p(y_{1:d}\mid\params), \quad d = 0, \ldots, T,
	\end{align*}
	and density tempering
	\begin{align*}
		\pi_d(\params_d) \ \propto \ p(\params)p(y_{1:T}\mid\params)^{g_d}, \quad 0 = g_0 < \cdots < g_{D} = 1.
	\end{align*}
	Likewise, a common choice for the mutation and backward kernels is to use $R$ applications of an MCMC kernel $\mathcal{K}$ with $\pi_d(\params_d)$ as its invariant distribution, and to set $L(\params_{d}, \params_{d-1}) = \gamma_d(\params_{d-1})K(\params_{d-1}, \params_{d})\slash \gamma_d(\params_{d})$ \citep{Chopin2002,DelMoral2006}. The weights in Step \ref{enum:step3} then simplify to 
	\begin{align}
		w_d^n = N_{\theta}^{-1}\cdot\frac{\gamma_d(\params_{d-1}^{n})}{\gamma_{d-1}(\params^n_{d-1})}, \quad W_d^n = \frac{w_d^n}{\sum_{i=1}^{N_{\theta}}w_d^i}, \label{eqn:smc_weights}
	\end{align}
	where $N_{\theta}$ is the number of parameter particles. As the weights are independent of the mutated particles, i.e.\ $w_d^n$ is independent of $\params^n_{d}$, the reweighting step (Step \ref{enum:step3}) is completed first, before the particles are resampled and mutated (Steps \ref{enum:step1} and \ref{enum:step2}). 
	
	\subsubsection{Particle Filters} \label{sec:pf}	
	If $\params$ is fixed and the parameters of interest are $z_{1:D} = x_{1:T}$, then SMC can be implemented by setting $\pi_T(x_{1:T}) = p(x_{1:T} \mid y_{1:T}, \params)$ and $\pi_1(x_1) = \mu(x_1\mid\params)$. For ease of exposition, we assume that $\pi_1(x_1)$ is the initial distribution (instead of $\pi_0(\cdot)$) and $d = 1, \ldots, D$ where $D = T$. Using data annealing, the intermediate distributions are
	\begin{align*}
		\pi_d(x_{1:d}) := \frac{p(x_{1:d}, y_{1:d} \mid \params)}{p(y_{1:d}\mid\params)} \ \propto \ \mu(x_1\mid\params)\prod_{i=2}^{d}{f(x_i \mid x_{i-1}, \params)}\prod_{i=1}^{d}{g(y_i\mid x_i, \params)}.
	\end{align*}
	The simplest particle filter is the bootstrap filter of \citet{Gordon1993}. In this case, the mutation kernel is the transition density and the backward kernel is 1, giving the weights in Step \ref{enum:step3} as 
	\begin{align*}
		w_d^n = W_{d-1}^n g(y_d\mid x_d^n, \params), \quad W_d^n = \frac{w_d^n}{\sum_{i=1}^{N_{x}}w_d^i}.
	\end{align*}
	Notably, for this particle filter, the right-hand side of \eqref{eqn:llest} gives an unbiased estimate of the likelihood $p(y_{1:T}\mid \params)$.
	
	Where the standard particle filter can be used to sample from $p(x_{1:T} \mid y_{1:T}, \params)$, the conditional particle filter (CPF) can be used to sample a latent state trajectory from its full conditional distribution \citep{Andrieu2010}. The CPF is identical to the standard particle filter, except that a single trajectory is taken as input and fixed throughout the iterations. Practically, this means that only $N_x - 1$ particles, i.e.\ all particles except the fixed or reference particle, are resampled and mutated in Steps \ref{enum:step1} and \ref{enum:step2}. Step \ref{enum:step3} remains unchanged.
	
	Once the CPF is run, a single trajectory, $x_{1:T}^{k_{1:T}}$, can be drawn from the resulting set of $N_x$ trajectories. Note that it is possible for the reference trajectory to be resampled. A naive approach samples $x_T^{k_T}$ from its full conditional distribution $\pi_T(\cdot)$, i.e.\ sample $x_T^{k_T}$ from $x_{T}^{1:N_x}$ using the normalized weights at time $d=T$, and then trace the ancestry of $x_T^{k_T}$ backward to $d=1$. This approach is referred to as ancestral tracing, and it is generally very sensitive to path degeneracy in the particle filter \citep{Lindsten2012}. An alternative method is backward sampling \citep{Whiteley2010,Lindsten2012}, which also draws $x_T^{k_T}$ from its full conditional distribution $\pi_T(\cdot)$, but instead of tracing the ancestry, $k_{1:T}$ is obtained using backward simulation. The weights, conditional on $x_T^{k_T}$, are calculated for $d = T-1, \ldots, 1$ as 
	\begin{align*}
		W_{d\mid T}^n = \frac{W_d^nf(x_{d+1}^{k_{d+1}}\mid x_d^n)}{\sum_{j = 1}^{N_x}W_d^jf(x_{d+1}^{k_{d+1}}\mid x_d^j)}, \quad n = 1, \ldots, N_x.
	\end{align*}
	At each iteration $d = T-1, \ldots, 1$, the index $k_d$ is drawn according to $W_{d\mid T}^{1:N_x}$. This approach greatly mitigates the effect of path degeneracy on the sampled trajectory. The standard and conditional particle filters are fundamental to SMC$^2$, as described in the next section. 
	
	\subsection{SMC\texorpdfstring{$^2$}{2}} \label{sec:smc2}	
	SMC$^2$ extends SMC for static models by either replacing the intractable likelihood with an unbiased particle filter estimator, or by using a conditional particle filter with backward sampling to mutate the latent state trajectories. These correspond to using a PMMH or PG algorithm to mutate the particles. Specific implementations of a PMMH and a PG kernel are given in Section \ref{sec:examples}. SMC$^2$ differs from the SMC method described in Section \ref{sec:smc_static} by the particular mutation kernel used and the form of the sequence of targets $\pi_d(\params_d), d = 0,\ldots, D$. The reweighting step remains unchanged, with the weights given in \eqref{eqn:smc_weights}. 
	
	Our aim is to switch between PMMH and PG when mutating the particles. For this to be feasible, the same distribution must be targeted by both methods at any given iteration. As with standard SMC for static models (see Section \ref{sec:smc_static}), the sequence of distributions can be constructed using data annealing or density tempering.
	
	\subsubsection{Data Annealing Targets} \label{sec:da_targets}
	Data annealing SMC$^2$ constructs the sequence of distributions by processing a new observation at each iteration. The $d$th unnormalised target in a data annealing sequence of distributions when using PMMH is then
	\begin{equation}
		\begin{aligned}
		\gamma_d(\params, x_{1:d}^{1:N_x}, y_{1:d}) \ &= \ p(\params)\widehat{p_{N_x}}(y_{1:d}\mid\params, x_{1:d}^{1:N_x})\psi(x_{1:d}^{1:N_x}\mid \params, y_{1:d}) \\
		&= \ p(\params)\widehat{p_{N_x}}(y_{1:d}, x_{1:d}^{1:N_x}\mid\params),
	\end{aligned}
	\label{eqn:da_pmmh_target}
	\end{equation}
	where $\psi(x_{1:d}^{1:N_x}\mid\cdot)$ is the proposal distribution for the latent state trajectories drawn using a particle filter, and $\widehat{p_{N_x}}(y_{1:d}, x_{1:d}^{1:N_x}\mid\params)$ is the particle filter estimate of the likelihood. The particle filter itself targets $p(x_{1:d}\mid y_{1:d}, \params)$. 
	
	Likewise, the $d$th unnormalised target in a data annealing sequence of distributions when using PG is 
	\begin{equation}
		\begin{aligned}
			&\gamma_d(\params, {k_{1:d}}, x_{1:d}^{k_{1:d}}, x_{1:d}^{1:N_x, -{k_{1:d}}}, y_{1:d}) \\ 
			&= \ p(\params)p(x_{1:d}^{k_{1:d}}\mid \params)p(y_{1:d}\mid\params, x_{1:d}^{k_{1:d}}) \psi_d(x_{1:d}^{1:N_x, -{k_{1:d}}}\mid\params, y_{1:d}, x_{1:d}^{k_{1:d}}, {k_{1:d}}) \\
			&\qquad \times\psi_d({k_{1:d}}\mid\params, y_{1:d}, x_{1:d}^{k_{1:d}}, x_{1:d}^{1:N_x, -{k_{1:d}}}) \\
			&= \ p(\params)p(x_{1:d}^{k_{1:d}}\mid \params)p(y_{1:d}\mid\params, x_{1:d}^{k_{1:d}}) \psi_d(x_{1:d}^{1:N_x, -{k_{1:d}}}, {k_{1:d}}\mid\params, y_{1:d}, x_{1:d}^{k_{1:d}}),
		\end{aligned}
		\label{eqn:da_pg_target}
	\end{equation}
	where $\psi_d(x_{1:d}^{1:N_x, -{k_{1:d}}}\mid\cdot)$ is the distribution of the latent state trajectories drawn using the conditional particle filter, and $\psi_d({k_{1:d}}\mid\cdot)$ is the distribution of the indices\slash trajectory of the invariant path, obtained e.g.\ using backward sampling (see Section \ref{sec:pf}). The target distribution of the conditional particle filter is $p(x_{1:d}\mid y_{1:d}, \params)$. 
 
 	The marginal density of \eqref{eqn:da_pmmh_target} and \eqref{eqn:da_pg_target} is $\pi_d(\params\mid y_{1:d}) \ \propto \ p(\params)p(y_{1:d}\mid\params)$ for $d = 0,\ldots, T$.
	
	\subsubsection{Density Tempering Targets} \label{sec:dt_targets}
	Density tempering SMC$^2$ constructs the sequence of distributions by raising the likelihood term to a power $g_d$, where $0 = g_0 < \ldots < g_D = 1$. The $d$th unnormalised target in a density tempering sequence of targets when using PMMH is then
	\begin{align}
		\gamma_d(\params, x_{1:T}^{1:N_x}, y_{1:T}) \ = \ p(\params)\left[\widehat{p_{N_x}}(y_{1:T}\mid\params, x_{1:T}^{1:N_x}) \right]^{g_d}\psi(x_{1:T}^{1:N_x}\mid \params, y_{1:T}), \label{eqn:dt_pmmh_target}
	\end{align}
	where $\psi(x_{1:T}^{1:N_x}\mid\cdot)$ is the proposal distribution for the latent state trajectories drawn using the particle filter. The target distribution of the particle filter is $p(x_{1:T}\mid y_{1:T}, \params)$.
	
	Similarly, the $d$th unnormalised target in a density tempering sequence of targets when using PG is
	\begin{equation}
		\begin{aligned}
			&\gamma_d(\params, {k_{1:T}}, x_{1:T}^{k_{1:T}}, x_{1:T}^{1:N_x, -{k_{1:T}}}, y_{1:T}) \\ &= \ p(\params)p(x_{1:T}^{k_{1:T}}\mid \params)\left[p(y_{1:T}\mid\params, x_{1:T}^{k_{1:T}}) \right]^{g_d} \psi_d(x_{1:T}^{1:N_x, -{k_{1:T}}}\mid\params, y_{1:T}, x_{1:T}^{k_{1:T}}, {k_{1:T}}) \\ &\qquad \times \psi_d({k_{1:T}}\mid\params, y_{1:T}, x_{1:T}^{k_{1:T}}, x_{1:T}^{1:N_x, -{k_{1:T}}}) \\
			&= \ p(\params)p(x_{1:T}^{k_{1:T}}\mid \params)\left[p(y_{1:T}\mid\params, x_{1:T}^{k_{1:T}}) \right]^{g_d} \psi_d(x_{1:T}^{1:N_x, -{k_{1:T}}}, {k_{1:T}}\mid\params, y_{1:T}, x_{1:T}^{k_{1:T}}), \label{eqn:dt_pg_target}
		\end{aligned}
	\end{equation}
	where $\psi_d(x_{1:T}^{1:N_x, -{k_{1:T}}}\mid\cdot)$ is the distribution of the latent state trajectories drawn using the conditional particle filter, and $\psi_d({k_{1:T}}\mid\cdot)$ is the distribution of the indices\slash trajectory of the invariant path. The target distribution of the conditional particle filter is $p(x_{1:T}\mid \params)\left[p(y_{1:T}\mid\params, x_{1:T}) \right]^{g_d}$. As a result, both $\psi_d(x_{1:T}^{1:N_x, -{k_{1:T}}}\mid\cdot)$ and $\psi_d({k_{1:T}}\mid\cdot)$ depend on $g_d$. 
	
	The marginal distribution for PMMH \eqref{eqn:dt_pmmh_target} is 
	\begin{align*}
		\pi_d(\params\mid y_{1:T}) \ \propto \ p(\params)\int{\left[\widehat{p_{N_x}}(y_{1:T}\mid\params, x_{1:T}^{1:N_x}) \right]^{g_d}\psi(x_{1:T}^{1:N_x}\mid \params, y_{1:T})}dx_{1:T}^{1:N_x},
	\end{align*}
	and the marginal distribution for PG \eqref{eqn:dt_pg_target} is 
	\begin{align*}
		\pi_d(\params\mid y_{1:T}) \ &\propto \ p(\params) \int p(x_{1:T}^{k_{1:T}}\mid \params)\left[p(y_{1:T}\mid\params, x_{1:T}^{k_{1:T}}) \right]^{g_d} \\ 
			&\qquad \times \psi_d(x_{1:T}^{1:N_x, -{k_{1:T}}}, {k_{1:T}}\mid\params, y_{1:T}, x_{1:T}^{k_{1:T}}) d{\{k_{1:T},x_{1:T}^{1:N_x}\}} \\
		&\propto \ p(\params)\int{p(x_{1:T}\mid \params)\left[p(y_{1:T}\mid\params, x_{1:T}) \right]^{g_d}}dx_{1:T}.
	\end{align*}
	Clearly, the distributions defined in \eqref{eqn:dt_pmmh_target} and \eqref{eqn:dt_pg_target} do not have the same marginal distributions unless $g_d = 0$ or $g_d = 1$, i.e.\ when $d = 0$ and $d = D$. 

	As an alternative to \eqref{eqn:dt_pmmh_target}, we consider
	\begin{align}
		\gamma_d(\params, x_{1:T}^{1:N_x}, y_{1:T}) \ = \ p(\params)\widehat{p_{N_x, d}}(y_{1:T}\mid\params, x_{1:T}^{1:N_x}) \psi_d(x_{1:T}^{1:N_x}\mid \params, y_{1:T}). \label{eqn:dt_pmmh_target_v2}
	\end{align} 
	Here, the particle filter targets $p(x_{1:T}\mid \params)$$\left[p(y_{1:T}\mid\params, x_{1:T}) \right]^{g_d}$, so $\widehat{p_{N_x, d}}(y_{1:T}\mid\params, x_{1:T}^{1:N_x})$ $\psi_d(x_{1:T}^{1:N_x}\mid \params, y_{1:T})$ is an unbiased estimate of 
	\begin{align*}
		\int{p(x_{1:T}\mid \params)\left[p(y_{1:T}\mid\params, x_{1:T}) \right]^{g_d}}dx_{1:T}.
	\end{align*}
	Hence, \eqref{eqn:dt_pg_target} and \eqref{eqn:dt_pmmh_target_v2} have the same marginal distributions for $0 \le g_d \le 1$. Note that the sequence defined by \eqref{eqn:dt_pmmh_target_v2} introduces extra variability in the reweighting step as the likelihood needs to be re-estimated whenever the tempering parameter changes. This also complicates adaptation of the tempering schedule $g_0, g_1, \cdots, g_D$. Due to these difficulties, we only consider \eqref{eqn:dt_pmmh_target_v2} when mutating the particles, and do not use it in the reweighting step. See Section \ref{sec:adaptation} for more details.
	
	\section{Switching Between Particle MCMC Kernels} \label{sec:switching_mechanics}
	This section describes the mechanics of switching between PMMH and PG kernels in SMC$^2$. Mathematical justification is given in Appendix \ref{app:kswitch}. Switching between particle MCMC kernels is essentially the same as switching from one sequence of targets to another. This is done within the mutation step only (Step \ref{enum:step3}) and does not change the reweighting or resampling steps (Steps \ref{enum:step1} and \ref{enum:step2}). For data annealing, the kernels considered are a PMMH algorithm targeting \eqref{eqn:da_pmmh_target} and a PG algorithm targeting \eqref{eqn:da_pg_target}. Likewise for density tempering, the kernels considered are a PMMH algorithm targeting \eqref{eqn:dt_pmmh_target_v2} and a PG algorithm targeting \eqref{eqn:dt_pg_target}. We denote the number of state particles used for PMMH and PG as $N_{x}^{\textrm{PMMH}}$ and $N_{x}^{\textrm{PG}}$ respectively. This section also shows how to switch between two PMMH or two PG kernels, i.e.\ where a different number of state particles is used.
	
	\subsection{Data Annealing}
	\begin{itemize}[labelindent=1pt,leftmargin=!, labelwidth=\widthof{aPMMH$\rightarrow$PMMHa}]
		\item[(PMMH $\rightarrow$ PG)] To switch from \eqref{eqn:da_pmmh_target} with $N_{x}^{\textrm{PMMH}}$ state particles to \eqref{eqn:da_pg_target} with $N_{x}^{\textrm{PG}}$ state particles, a particle filter targeting $p(x_{1:d}\mid y_{1:d}, \params)$ is run with $N_{x}^{\textrm{PG}}$ to obtain $\{x_{1:d}^{n}, W_{1:d}^{n}\}_{n = 1}^{N_{x}^{\textrm{PG}}}$. A new invariant path is drawn from $\{x_{1:d}^{n}, W_{1:d}^{n}\}_{n = 1}^{N_{x}^{\textrm{PG}}}$ using backward sampling as described in Section \ref{sec:pf}. 
		\item[(PG $\rightarrow$ PMMH)] To switch from \eqref{eqn:da_pg_target} with $N_{x}^{\textrm{PG}}$ state particles to \eqref{eqn:da_pmmh_target} with $N_{x}^{\textrm{PMMH}}$ state particles, a particle filter targeting $p(x_{1:d}\mid y_{1:d}, \params)$ is run with $N_{x}^{\textrm{PMMH}}$ state particles to obtain the likelihood estimate $\widehat{p_{N_{x}^{\textrm{PMMH}}}}(y_{1:d}$, $x_{1:d}^{1:N_{x}^{\textrm{PMMH}}}\mid\params)$. 
		\item[(PMMH $\rightarrow$ PMMH)] To switch from \eqref{eqn:da_pmmh_target} with $N_{x}^{\textrm{PMMH}, 1}$ state particles to \eqref{eqn:da_pmmh_target} with $N_{x}^{\textrm{PMMH}, 2}$ state particles, a particle filter targeting $p(x_{1:d}\mid y_{1:d}, \params)$ is run with $N_{x}^{\textrm{PMMH}, 2}$ state particles to obtain the likelihood estimate \\$\widehat{p_{N_{x}^{\textrm{PMMH}, 2}}}(y_{1:d}, x_{1:d}^{1:N_{x}^{\textrm{PMMH}, 2}}\mid\params)$. 
		\item[(PG $\rightarrow$ PG)] To switch from \eqref{eqn:da_pg_target} with $N_{x}^{\textrm{PG}, 1}$ state particles to \eqref{eqn:da_pg_target} with $N_{x}^{\textrm{PG}, 2}$ state particles, a conditional particle filter targeting $p(x_{1:d}\mid y_{1:d}, \params, x_{1:d}^{k_{1:d}})$ is run with $N_{x}^{\textrm{PG}, 2}$ state particles to obtain $\{x_{1:d}^{n}, W_{1:d}^{n}\}_{n = 1}^{N_{x}^{\textrm{PG}, 2}}$. A new invariant path is drawn from $\{x_{1:d}^{n}, W_{1:d}^{n}\}_{n = 1}^{N_{x}^{\textrm{PG}, 2}}$ using backward sampling as described in Section \ref{sec:pf}.
	\end{itemize}

	\subsection{Density Tempering}
	\begin{itemize}[labelindent=1pt,leftmargin=!, labelwidth=\widthof{aPMMH$\rightarrow$PMMHa}]
		\item[(PMMH $\rightarrow$ PG)] To switch from \eqref{eqn:dt_pmmh_target_v2} with $N_{x}^{\textrm{PMMH}}$ state particles to \eqref{eqn:dt_pg_target} with $N_{x}^{\textrm{PG}}$ state particles, a particle filter targeting $p(x_{1:T}\mid \params)\left[p(y_{1:T}\mid\params, x_{1:T}) \right]^{g_d}$ is run with $N_{x}^{\textrm{PG}}$ state particles to obtain $\{x_{1:T}^{n}, W_{1:T}^{n}\}_{n = 1}^{N_{x}^{\textrm{PG}}}$. A new invariant path is drawn from $\{x_{1:T}^{n}, W_{1:T}^{n}\}_{n = 1}^{N_{x}^{\textrm{PG}}}$ using backward sampling as described in Section \ref{sec:pf}.
		\item[(PG $\rightarrow$ PMMH)] To switch from \eqref{eqn:dt_pg_target} with $N_{x}^{\textrm{PG}}$ state particles to \eqref{eqn:dt_pmmh_target_v2} with $N_{x}^{\textrm{PMMH}}$ state particles, a particle filter targeting $p(x_{1:T}\mid \params)\left[p(y_{1:T}\mid\params, x_{1:T}) \right]^{g_d}$ is run with $N_{x}^{\textrm{PMMH}}$ state particles to obtain the likelihood estimate $\widehat{p_{N_{x}^{\textrm{PMMH}},d}}(y_{1:T}, x_{1:T}^{1:N_{x}^{\textrm{PMMH}}}\mid\params)$.
		\item[(PMMH $\rightarrow$ PMMH)] To switch from \eqref{eqn:dt_pmmh_target_v2} with $N_{x}^{\textrm{PMMH}, 1}$ state particles to \eqref{eqn:dt_pmmh_target_v2} with $N_{x}^{\textrm{PMMH}, 2}$ state particles, a particle filter targeting $p(x_{1:T}\mid \params)$ $\left[p(y_{1:T}\right.$$\mid$ $\params$$, $$x_{1:T})$$\left.\right]^{g_d}$ is run with $N_{x}^{\textrm{PMMH}, 2}$ state particles to obtain the likelihood estimate $\widehat{p_{N_{x}^{\textrm{PMMH}, 2}, d}}(y_{1:T}, x_{1:T}^{1:N_{x}^{\textrm{PMMH}, 2}}\mid\params)$.
		\item[(PG $\rightarrow$ PG)] To switch from \eqref{eqn:dt_pg_target} with $N_{x}^{\textrm{PG}, 1}$ state particles to \eqref{eqn:dt_pg_target} with $N_{x}^{\textrm{PG}, 2}$ state particles, a conditional particle filter targeting $p(x_{1:d}\mid y_{1:d}, \params)$ is run with $N_{x}^{\textrm{PG}, 2}$ state particles to obtain $\{x_{1:d}^{n}, W_{1:d}^{n}\}_{n = 1}^{N_{x}^{\textrm{PG}, 2}}$. A new invariant path is drawn from $\{x_{1:d}^{n}, W_{1:d}^{n}\}_{n = 1}^{N_{x}^{\textrm{PG}, 2}}$ using backward sampling as described in Section \ref{sec:pf}.
	\end{itemize}
	
	\section{Adaptation} \label{sec:adaptation}
	This section focuses on when to switch between a PMMH and PG kernel. At each iteration, the candidate kernels are scored using the squared jumping distance of the particles as a measure of statistical efficiency and the number of state particles as a measure of the computation cost. The kernel with the highest score is used to mutate the particles. 
	
	\subsection{Squared Jumping Distance}
	The expected squared jumping distance \citep[ESJD; ][]{Pasarica2010} is a useful metric for the performance of an MCMC kernel in SMC \citep{Fearnhead2013,Salomone2018,Bon2021} and SMC$^2$ \citep{Botha2023} as it takes both the acceptance rate and the jumping distance of the particles into account. The ESJD for the $i$th parameter particle, for $i = 1, \ldots, N_{\theta}$, at iteration $d$ of the SMC algorithm can be estimated as
	\begin{align*}
		\widehat{\textrm{ESJD}} &= \frac{1}{N_\theta}\sum_{i=1}^{N_\theta}{\widehat{H}(\params^i_{d}, \params^*_{d})}, \\
		\widehat{H}(\params^i_{d}, \params^*_{d}) &= \norm{\params^*_{d} - \params^i_{d}}^2\alpha(\params^i_{d}, \params^*_{d}) \\
		&= (\params^i_{d} - \params^*_{d})^\top\widehat\Sigma_d^{-1}(\params^i_{d}-\params^*_{d})\alpha(\params^i_{d}, \params^*_{d}),
	\end{align*}
	where $\params^{i}_{d}$ is the current value of the $i$th parameter particle, $\params^*_{d}$ is the proposed value for the $i$th parameter particle, $\norm{\params^*_{d} - \params^i_{d}}$ is the estimated Mahalanobis distance between $\params^{i}_{d}$ and $\params^*_{d}$, $\widehat\Sigma_d$ is the covariance of the current parameter particle set $\params_{d}^{1:N_{\theta}}$, and $\alpha(\params^i_{d}, \params^*_{d})$ is the Metropolis-Hastings acceptance probability. 
	
	It is unclear, however, how to estimate the ESJD when the parameters are updated singly or in blocks, or how to estimate it over multiple MCMC iterations. As a simplified but more versatile measure, we consider the squared jumping distance (SJD) instead of the ESJD. Advantageously, the SJD can be calculated for any form of update and over any number of iterations. Denote the number of MCMC iterations as $K$, the original value of the $i$th parameter particle as $\params^{i}_{d, 0}$, and its value after $K$ particle  MCMC iterations as $\params^{i}_{d, K}$. Then, the SJD between $\params_{d, 0}^{i}$ and $\params_{d, K}^{i}$ for $i = 1, \ldots, N_{\theta}$ is
	\begin{align}
		\textrm{SJD}_{d, K}^{i} = (\params_{d, 0}^{i} - \params_{d, K}^{i})^\top\widehat\Sigma_d^{-1}(\params_{d, 0}^{i}-\params_{d, K}^{i}). \label{eqn:sjd}
	\end{align}

	While the SJD is typically calculated as a joint metric over the parameters, it can also be calculated for each parameter as 
	\begin{align}
		\textrm{\textbf{pSJD}}_{d, K} &= \frac{1}{N_{\theta}}\sum_{i=1}^{N_{\theta}}{v_{d, K}^i\odot v_{d, K}^i}, \quad v_{d, K}^i = (\params_{d, 0}^{i} - \params_{d, K}^{i})^\top\widehat\Sigma_d^{-1\slash 2},
		\label{eqn:sjd_breakdown}
	\end{align}
	where $\odot$ denotes element-wise multiplication, and $\textrm{\textbf{pSJD}}_{d, K}$ is a vector of the same length as $\params$. Each element in $\textrm{\textbf{pSJD}}_{d, K}$ gives the average SJD for the corresponding parameter in $\params$ after $K$ iterations of the particle MCMC algorithm. This is useful when adapting the number of particle MCMC repeats, as it ensures that a minimum SJD can be targeted for each parameter. Note that the sum of the elements of $\textrm{\textbf{pSJD}}_{d, K}$ gives the mean of the SJD in \eqref{eqn:sjd} with respect to the parameter particles, i.e.\ $\textbf{1}^{\top}\textrm{\textbf{pSJD}}_{d, K} = \Nt^{-1}\sum_{i=1}^{\Nt}\textrm{SJD}_{d, K}^{i}$, where $\textbf{1}$ is a vector of ones of the same length as $\textrm{\textbf{pSJD}}_{d, K}$.
	
	We calculate the covariance of the particles, $\Sigma_d$, before the resampling step using the weighted covariance formula of \citet{Price1971}. This covariance is used for the particle MCMC proposal functions and to calculate the SJD throughout the algorithm. Fractional powers of matrices are calculated using the method outlined in \citet{Higham2013}.
	
	\subsection{Choosing the best mutation kernel}
	We consider two candidate particle MCMC kernels, a PMMH kernel and a PG kernel. One of these will be the default kernel, denoted $\mathcal{K}_{\textrm{def}}$, which is both the initial kernel used during the mutation step and also the one that defines the sequence of targets to be used in the reweighting step. The remaining kernel is referred to as the alternate kernel, denoted $\mathcal{K}_{\textrm{alt}}$. As an example, if $\mathcal{K}_{\textrm{def}}$ is a PMMH kernel targeting \eqref{eqn:da_pmmh_target}, then $\mathcal{K}_{\textrm{alt}}$ is a PG kernel targeting \eqref{eqn:da_pg_target}. In this case, $\mathcal{K}_{\textrm{def}}$ or $\mathcal{K}_{\textrm{alt}}$ can be used to mutate the particles, but the SMC weights in \eqref{eqn:smc_weights} are always calculated using \eqref{eqn:da_pmmh_target}. 
	
	Our proposed approach proceeds as follows. At the start of the mutation step, a score is calculated for $\mathcal{K}_{\textrm{def}}$. The particle MCMC kernel is then switched to $\mathcal{K}_{\textrm{alt}}$ and a score is calculated for $\mathcal{K}_{\textrm{alt}}$. If $\mathcal{K}_{\textrm{def}}$ has the highest score, the kernel is switched back to $\mathcal{K}_{\textrm{def}}$ and the mutation is completed. If however, $\mathcal{K}_{\textrm{alt}}$ has the higher score, the mutation is first completed with $\mathcal{K}_{\textrm{alt}}$ and then the kernel is switched back to $\mathcal{K}_{\textrm{def}}$ for reweighting. The adaptive move step is given in Algorithm \ref{alg:adapt_mutation}.
	
	The score for a given particle MCMC kernel $\mathcal{K}$ is calculated as $\nicefrac{m_{d, K}^{\mathcal{K}}}{N_x^{\mathcal{K}}}$, where $m_{d, K}^{\mathcal{K}} = \min{(\textrm{\textbf{pSJD}}_{d, K}^{\mathcal{K}})}$ is the minimum SJD over the parameters, and $N_x^{\mathcal{K}}$ is the number of state particles used for $\mathcal{K}$. Denote the kernel with the highest score as $\mathcal{K}_{\textrm{best}}$. After both kernels have been tested, the remaining number of iterations is calculated as
	\begin{align}
		R_{\textrm{rem}} = \ceil{\frac{\textrm{SJD}_{\textrm{target}} - \min{(\textrm{\textbf{pSJD}}_{d, K}^{\mathcal{K}_{\textrm{def}}} + \textrm{\textbf{pSJD}}_{d, K}^{\mathcal{K}_{\textrm{alt}}})}}{m_{d, K}^{\mathcal{K}_{\textrm{best}}}\slash K}}, \label{eqn:Rrem}
	\end{align}
	where $\textrm{SJD}_{\textrm{target}}$ is the minimum target SJD for each parameter and $K$ is the number of iterations used to estimate the SJD. The numerator gives the new SJD target after accounting for the movement from testing both kernels, and the denominator gives the average minimum SJD (i.e.\ the average over the $K$ particle MCMC repeats) for the kernel with the highest score.

	The particle MCMC kernels do not necessarily need to be tested at every iteration. Another option is to test both kernels in the early stages of the algorithm (e.g.\ in the first 5 iterations), and thereafter only at certain points specified by a lag parameter. The latter gives the number of iterations before $\mathcal{K}_{\textrm{alt}}$ should be tested again, and is calculated as the score for $\mathcal{K}_{\textrm{def}}$ divided by the score for $\mathcal{K}_{\textrm{alt}}$. For example, if the score for $\mathcal{K}_{\textrm{def}}$ is $2$ and the score for $\mathcal{K}_{\textrm{alt}}$ is anywhere between 1 and 2, then $\mathcal{K}_{\textrm{alt}}$ will be tested after $2$ iterations from the current iteration. The aim here is to reduce the number of times $\mathcal{K}_{\textrm{alt}}$ is tested if $\mathcal{K}_{\textrm{def}}$ consistently outperforms $\mathcal{K}_{\textrm{alt}}$.
		
	The only parameters that must be set for this framework are $K$ and $\textrm{SJD}_{\textrm{target}}$. In all our numerical experiments we used $K = 5$ and we adapt $\textrm{SJD}_{\textrm{target}}$ at each iteration prior to the resampling and mutation step. Specifically, we set 
	\begin{align*}
		\textrm{SJD}_{\textrm{target}} = 4\sum_{i=1}^{\Nt}{W_{d}^{i}(\params_{d}^{i} - \vect{\mu}_{d})^\top\widehat\Sigma_d^{-1}(\params_{d}^{i} - \vect{\mu}_{d})},
	\end{align*}
	where $W_{d}^{i}$ is the normalised weight associated with $\params_{d}^{i}$, $\widehat\Sigma_d$ is the weighted covariance of $\params_{d}^{1:\Nt}$, and $\vect{\mu}_{d}$ is the weighted mean of $\params_{d}^{1:\Nt}$. Recall, after the reweighting step, the weighted samples form an empirical approximation of $\pi_d(\params_d)$. Assuming this approximation is good, 4 times the weighted SJD of the particles to their mean should allow for sufficient exploration of the current distribution in the mutation step. In practice, we find that this is the case.
	
	\begin{algorithm}[htp]
		\begin{adjustwidth}{\algorithmicindent}{}
			\textbf{Input: } current set of particles $\params_{d}^{1:N_{\theta}}$, particle MCMC kernel $\mathcal{K}$ \\
			\textbf{Output: } mutated set of particles $\params_{d}^{1:N_{\theta}}$, SJD over the parameters $\textrm{\textbf{pSJD}}_{d, K}^{\mathcal{K}}$ 
		\end{adjustwidth}
		\begin{algorithmic}			
			\State Set $\params_{d, 0}^{1:N_{\theta}} = \params_{d}^{1:N_{\theta}}$
			\For{$k = 1$ to $K$}
			\State $\params_{d, k}^{1:N_{\theta}} \sim \mathcal{K}(\params_{d, k-1}^{1:N_{\theta}}, \params_{d, k}^{1:N_{\theta}})$, i.e.\ using Algorithm \ref{alg:pmmh_kernel} or Algorithm \ref{alg:pg_kernel}
			\EndFor
			\State Calculate $\textrm{\textbf{pSJD}}_{d, K}^{\mathcal{K}}$ using \eqref{eqn:sjd_breakdown}
			\State Set $\params_{d}^{1:N_{\theta}} = \params_{d, K}^{1:N_{\theta}}$
		\end{algorithmic}
		\caption{Test the particle MCMC kernel}
		\label{alg:test_mutation_kernel}
	\end{algorithm}
	
	\begin{algorithm}[htp]
		\begin{adjustwidth}{\algorithmicindent}{}
			\textbf{Input: } current set of particles $\params_{d}^{1:N_{\theta}}$\\
			\textbf{Output: } mutated set of particles $\params_{d}^{1:N_{\theta}}$
		\end{adjustwidth}
		\begin{algorithmic}
			\LineComment{Test the default kernel}
			\vspace{0.5em}			
			\State Run Algorithm \ref{alg:test_mutation_kernel} to obtain $\textrm{\textbf{pSJD}}_{d, K}^{\mathcal{K}_{\textrm{def}}}$. Calculate $m_{d, K}^{\mathcal{K}_{\textrm{def}}} = \min{(\textrm{\textbf{pSJD}}_{d, K}^{\mathcal{K}_{\textrm{def}}})}$.
			
			\vspace{0.5em}
			\LineComment{Test the alternate kernel}
			\State Switch to $\mathcal{K}_{\textrm{alt}}$ as discussed in Section \ref{sec:switching_mechanics}.
			\State Run Algorithm \ref{alg:test_mutation_kernel} to obtain $\textrm{\textbf{pSJD}}_{d, K}^{\mathcal{K}_{\textrm{alt}}}$. Calculate $m_{d, K}^{\mathcal{K}_{\textrm{alt}}} = \min{(\textrm{\textbf{pSJD}}_{d, K}^{\mathcal{K}_{\textrm{alt}}})}$.
			
			\vspace{0.5em}
			\LineComment{Complete the remaining mutation steps}
			\State Calculate the number of remaining MCMC iterations using Equation \eqref{eqn:Rrem}.
			\If{$\nicefrac{m_{d, K}^{\mathcal{K}_{\textrm{def}}}}{N_x^{\mathcal{K}_{\textrm{def}}}} \ge \nicefrac{m_{d, K}^{\mathcal{K}_{\textrm{alt}}}}{N_x^{\mathcal{K}_{\textrm{alt}}}}$}
				\State Switch back to $\mathcal{K}_{\textrm{def}}$
				\For{$r = 1$ to $R_{\textrm{rem}}$}
				\State Mutate the particles using $\mathcal{K}_{\textrm{def}}$, i.e.\ using Algorithm \ref{alg:pmmh_kernel} or Algorithm \ref{alg:pg_kernel}
				\EndFor
			\Else
				\For{$r = 1$ to $R_{\textrm{rem}}$}
				\State Mutate the particles using $\mathcal{K}_{\textrm{alt}}$, i.e.\ using Algorithm \ref{alg:pmmh_kernel} or Algorithm \ref{alg:pg_kernel}
				\EndFor
				\State Switch back to $\mathcal{K}_{\textrm{def}}$
			\EndIf
		\end{algorithmic}
		\caption{Adaptive move step}
		\label{alg:adapt_mutation}
	\end{algorithm}

	\section{Examples} \label{sec:examples}
	Sections \ref{sec:mutation_kernels}-\ref{sec:scores} describe the implementation of our numerical experiments, including the specific PMMH and PG kernels used, the setup of the experiments and how the performance of the methods is evaluated. Sections \ref{sec:bm}-\ref{sec:ar1} demonstrate the performance of our approach on a Brownian motion model, a flexible-allee logistic model, a stochastic volatility in mean model and an AR(1) model.
	
	\subsection{Particle MCMC Kernels} \label{sec:mutation_kernels}
	For ease of exposition, we use the notation for data annealing SMC$^2$ in this section. To apply to density tempering SMC$^2$, simply replace $x_{1:d}^{1:N_x}$ with $x_{1:T}^{1:N_x}$ and $k_{1:d}$ with $k_{1:T}$.
	
	The first particle MCMC kernel is described in Algorithm \ref{alg:pmmh_kernel} and uses a single iteration of PMMH to jointly update the parameters in $\params$. We use a random walk proposal with covariance given by $\epsilon_{\textrm{PMMH}}^2\Sigma_d$, where $\epsilon_{\textrm{PMMH}}^2 \le 1$ is a stepsize that ensures the acceptance rate does not fall below $0.07$ (following the general recommendation of \citealt{Sherlock2015} for PMMH), and $\Sigma_d$ is an estimate of the covariance of the current target. We find that the stepsize $\epsilon_{\textrm{PMMH}}^2$ is useful in the initial stages of the algorithm if the estimate of the target covariance is poor. Generally, however, $\epsilon_{\textrm{PMMH}}^2 = 1$. 
	
	The second particle MCMC kernel is based on PG and is described in Algorithm \ref{alg:pg_kernel}. For each parameter particle, we draw a new invariant path $x_{1:d}^{k_{1:d}}$ using a conditional particle filter with backward sampling as described in Section \ref{sec:pf}. Conditional on this trajectory, the parameters are updated in two blocks. Let $\params_{d} = \{\vect{\phi}_{d, 1}, \vect{\phi}_{d, 2}\}$. The first block of parameters $\vect{\phi}_{d, 1}$ contains all of the parameters in the transition density $f(x_t\mid x_{t-1}, \params_{d})$ and $\vect{\phi}_{d, 2}$ contains the remaining parameters, including those that occur only in the observation density $g(y_{t}\mid x_{t}, \params_{d})$. The transition and observation densities can then be written as $f(x_t\mid x_{t-1}, \vect{\phi}_{d, 1})$ and $g(y_{t}\mid x_{t}, \vect{\phi}_{d, 1}, \vect{\phi}_{d, 2})$ respectively. In our numerical experiments, we always have $g(y_{t}\mid x_{t}, \vect{\phi}_{d, 1}, \vect{\phi}_{d, 2}) = g(y_{t}\mid x_{t}, \vect{\phi}_{d, 2})$. To update each block of parameters, we use MALA \citep{Girolami2011} with proposal distributions given by 
	\begin{align*}
		q_1(\cdot\mid\vect{\phi}_{d, 1}, \vect{\phi}_{d, 2}) = \mathcal{N}(\cdot\mid \vect{\phi}_{d, 1} + \frac{1}{2}\epsilon_1^2 \Sigma_{d, 1} \nabla_{\vect{\phi}_{d, 1}}\log{\gamma_d(\vect{\phi}_{d, 1}, \vect{\phi}_{d, 2}, x_{1:d}^{k_{1:d}})}, \epsilon_1^2 \Sigma_{d, 1}),
	\end{align*}
	and 
	\begin{align*}
		q_2(\cdot\mid\vect{\phi}_{d, 1}, \vect{\phi}_{d, 2}) = \mathcal{N}(\cdot\mid \vect{\phi}_{d, 2} + \frac{1}{2}\epsilon_2^2 \Sigma_{d, 2} \nabla_{\vect{\phi}_{d, 2}}\log{\gamma_d(\vect{\phi}_{d, 1}, \vect{\phi}_{d, 2}, x_{1:d}^{k_{1:d}})}, \epsilon_2^2 \Sigma_{d, 2}).
	\end{align*}
	Here, $\epsilon_1^2$ and $\epsilon_2^2$ are the MALA stepsizes, $\Sigma_{d}$ is the covariance of the current set of parameter particles $\params_{d}^{1:N_{\theta}}$, $\Sigma_{d, 1}$ and $\Sigma_{d, 2}$ are submatrices of $\Sigma_{d}$ corresponding to the parameters in $\vect{\phi}_{d, 1}$ and $\vect{\phi}_{d, 2}$ respectively, and $\gamma_d(\vect{\phi}_{d, 1}, \vect{\phi}_{d, 2}, x_{1:d}^{k_{1:d}})$ is the current target as defined in \eqref{eqn:da_pg_target} or \eqref{eqn:dt_pg_target} (depending on whether data annealing or density tempering is used). 
	
	Since the updates of $\vect{\phi}_{d, 1}\mid x_{1:d}^{k_{1:d}}, \vect{\phi}_{d, 2}$ and $\vect{\phi}_{d, 2}\mid x_{1:d}^{k_{1:d}}, \vect{\phi}_{d, 1}$ are relatively inexpensive, they are repeated $K_{\textrm{MALA}}$ times conditional on the same invariant path. A separate stepsize is adapted for each parameter block to achieve an average acceptance rate of $0.574$ \citep{Roberts2001}.
	
	For both particle MCMC kernels, the stepsizes are adapted as
	\begin{align*}
		\epsilon^2 = \epsilon^2 \cdot \exp{\left(2\left(\frac{\textrm{ar}}{\textrm{ar}_{\textrm{target}}} - 1\right)\right)},
	\end{align*}
	where $\textrm{ar}$ is the average acceptance rate over the parameter particles, and $\textrm{ar}_{\textrm{target}}$ is the target acceptance rate. In essence, this approach scales the standard deviation of the proposal function by $\exp{\left({\textrm{ar}_{\textrm{target}}}^{-1}(\textrm{ar} -{\textrm{ar}_{\textrm{target}}})\right)}$. Recall, $\textrm{ar}_{\textrm{target}} = 0.07$ for PMMH and $\textrm{ar}_{\textrm{target}} = 0.574$ for PG.  If $\textrm{ar}\approx{\textrm{ar}_{\textrm{target}}}$, the scaling factor will be $1$. In practice, we find that this approach works well to keep the average acceptance rate at the targeted level. 
	
	The stepsizes for PMMH and PG, the weighted covariance of the parameter particles and the target SJD are calculated at each iteration just prior to the resampling step. The acceptance rate for a particle MCMC kernel is calculated as the average acceptance rate over the parameter particles and the particle MCMC repeats, that is, $\Nt$ and $R$, respectively.
	
	\begin{algorithm}[htp]
		\begin{adjustwidth}{\algorithmicindent}{}
			\textbf{Input: } current set of particles $\params_{d}^{1:N_{\theta}}$ \\
			\textbf{Output: } mutated set of particles $\params_{d}^{1:N_{\theta}}$ 
		\end{adjustwidth}
		\vspace{0.5em}
		\begin{algorithmic}		
			\vspace{0.5em}
			\For{$i = 1$ to $N_{\theta}$}
			\State Draw $\params^{*}\sim \mathcal{N}(\cdot\mid\params_{d}^{i},  \epsilon_{\textrm{PMMH}}^2\Sigma_d)$
			\vspace{0.5em}
			\State Run a particle filter to draw $x_{1:d}^{1:N_x^{\textrm{PMMH}}, *}$ 
			\vspace{0.5em}
			\State Calculate the acceptance probability
			\begin{align*}
				\alpha(\params_{d}^i, \params^{*}) = \min{\left(1, \frac{\gamma_d(\params^{*}, x_{1:d}^{1:N_x^{\textrm{PMMH}}, *})}{\gamma_d(\params_{d}^{i}, x_{1:d}^{1:N_x^{\textrm{PMMH}}})}
					\right)
					} 
			\end{align*}
			\vspace{0.5em}
			\State With probability $\alpha(\params_{d}^{i}, \params^{*})$, set $\params_{d}^{i} = \params^{*}$ and $x_{1:d}^{1:N_x^{\textrm{PMMH}}} = x_{1:d}^{1:N_x^{\textrm{PMMH}}, *}$
			\EndFor
		\end{algorithmic}
		\caption{Particle marginal Metropolis-Hastings kernel.}
		\label{alg:pmmh_kernel}
	\end{algorithm}
	
	\begin{algorithm}[htp]
		\begin{adjustwidth}{\algorithmicindent}{}
			\textbf{Input: } current set of particles $\{\params_{d}^{i}, x_{1:d}^{k_{1:d}^i, i}\}_{i=1}^{N_{\theta}}$\\
			\textbf{Output: } mutated set of particles $\{\params_{d}^{i}, x_{1:d}^{k_{1:d}^i, i}\}_{i=1}^{N_{\theta}}$ \\
			\textbf{Note: } for this algorithm, the parameters are divided into two blocks, $\params_{d}^{i} = \{\vect{\phi}_{d, 1}^{i}, \vect{\phi}_{d, 2}^{i}\}$.
		\end{adjustwidth}
		\begin{algorithmic}		
			\LineComment{Update the latent states}
			\For{$i = 1$ to $N_{\theta}$}
			\State Run a conditional particle filter to draw $x_{1:d}^{1:N_x^{\textrm{PG}}, -{k_{1:d}^i}, i}\mid x_{1:d}^{k_{1:d}^i, i}$ 
			\vspace{0.5em}
			\State Use backward sampling to draw $k_{1:d}^i\mid x_{1:d}^{1:N_x^{\textrm{PG}}, -{k_{1:d}^i}, i}, x_{1:d}^{k_{1:d}^i, i}$ as described in Section \ref{sec:pf}
			\EndFor
			\vspace{0.5em}
			\LineComment{Update the static parameters}
			\For{$i = 1$ to $N_{\params}$}
			\For{$m = 1$ to $K_{\textrm{MALA}}$}
			\LineCommentIndent{Update the parameters in block 1}
			\State Draw $\vect{\phi}_{d, 1}^{*} \sim q_1(\cdot\mid\vect{\phi}_{d, 1}^{i}, \vect{\phi}_{d, 2}^{i})$
			\vspace{0.5em}
			\State Calculate the acceptance probability
			\begin{align*}
				\alpha(\vect{\phi}_{d, 1}^{i}, \vect{\phi}_{d, 1}^{*}) = \min{\left(1, \frac{\gamma_d(\vect{\phi}_{d, 1}^{*}, \vect{\phi}_{d, 2}^{i}, x_{1:d}^{k_{1:d}^i, i})}{\gamma_d(\vect{\phi}_{d, 1}^{i}, \vect{\phi}_{d, 2}^{i}, x_{1:d}^{k_{1:d}^i, i})} 
					\frac{q_1(\vect{\phi}_{d, 1}^{i}\mid\vect{\phi}_{d, 1}^{*}, \vect{\phi}_{d, 2}^{i})}{q_1(\vect{\phi}_{d, 1}^{*}\mid\vect{\phi}_{d, 1}^{i}, \vect{\phi}_{d, 2}^{i})}\right)}
			\end{align*}
			\vspace{0.5em}
			\State With probability $\alpha(\vect{\phi}_{d, 1}^{i}, \vect{\phi}_{d, 1}^{*})$, set $\vect{\phi}_{d, 1}^{i} = \vect{\phi}_{d, 1}^{*}$ and update $\params_{d}^{i}$
			\vspace{0.5em}
			\LineComment{Update the parameters in block 2}
			\State Draw $\vect{\phi}_{d, 2}^{*} \sim q_2(\cdot\mid\vect{\phi}_{d, 1}^{i}, \vect{\phi}_{d, 2}^{i})$
			\vspace{0.5em}
			\State Calculate the acceptance probability
			\begin{align*}
				\alpha(\vect{\phi}_{d, 2}^{i}, \vect{\phi}_{d, 2}^{*}) = \min{\left(1, \frac{\gamma_d(\vect{\phi}_{d, 1}^{i}, \vect{\phi}_{d, 2}^{*}, x_{1:d}^{k_{1:d}^i, i})}{\gamma_d(\vect{\phi}_{d, 1}^{i}, \vect{\phi}_{d, 2}^{i}, x_{1:d}^{k_{1:d}^i, i})} 
					\frac{q_2(\vect{\phi}_{d, 2}^{i}\mid\vect{\phi}_{d, 1}^{i}, \vect{\phi}_{d, 2}^{*})}{q_2(\vect{\phi}_{d, 2}^{*}\mid\vect{\phi}_{d, 1}^{i}, \vect{\phi}_{d, 2}^{i})}\right)}
			\end{align*}
			\vspace{0.5em}
			\State With probability $\alpha(\vect{\phi}_{d, 2}^{i}, \vect{\phi}_{d, 2}^{*})$, set $\vect{\phi}_{d, 2}^{i} = \vect{\phi}_{d, 2}^{*}$ and update $\params_{d}^{i}$
			\EndFor
			\EndFor
		\end{algorithmic}
		\caption{Single mutation using particle Gibbs}
		\label{alg:pg_kernel}
	\end{algorithm}

	\subsection{Details of Numerical Experiments} \label{sec:setup}
	We evaluate our methods on a Brownian motion model, a flexible-allee logistic growth model, a stochastic volatility in mean model and an AR(1) model. 
	
	All code is implemented in MATLAB. Adaptive multinomial resampling is used for the particle filter, the conditional particle filter and data annealing SMC$^2$, i.e.\ resampling is done whenever the effective sample size (ESS) falls below $\nicefrac{N_x}{2}$ or $\nicefrac{N_{\theta}}{2}$. For targets constructed using density tempering, the tempering parameter is adapted at each iteration to give an ESS of $\nicefrac{N_{\theta}}{2}$. The number of state particles used for the PMMH kernels ($N_{x}^{\textrm{PMMH}}$) is set for each model so that the variance of the log of the likelihood estimator is around $1$ when evaluated at the true values (for simulated data) or the posterior mean from a pilot run of SMC$^2$ (for real data). For PG kernels, $N_{x}^{\textrm{PG}}$ is set to a fraction of $N_{x}^{\textrm{PMMH}}$, i.e.\ $N_{x}^{\textrm{PG}} = r\cdot N_{x}^{\textrm{PMMH}}$. We test $r = 1, 0.5, 0.2, 0.05$. Note, however, that our aim is not to select or determine the best performing $N_x^{\textrm{PG}}$ for particle Gibbs. Instead, we use the different values of $N_{x}^{\textrm{PG}}$ to show the performance of our method when the alternate (or default) kernel is more or less efficient than the remaining kernel.
		
	As mentioned in Section \ref{sec:adaptation}, we use $K = 5$ applications of a particle MCMC kernel to test its performance. We also set the number of MALA iterations conditioned on the same invariant path to $5$, $K_{\textrm{MALA}} = 5$. We test our switching methodology when the alternate kernel is always tested at the start of the mutation step, and also when a lag parameter is used. In both cases, the alternate kernel is always tested in the first $5$ iterations of the algorithm.
	
	In practice, we find that the variability of the reweighting step is high when PG is the default kernel, $r$ is less than $1$, and the kernels are allowed to switch. When switching from PMMH to PG, a new set of trajectories is drawn using a standard particle filter with $N_{x}^{\textrm{PG}} = r\cdot N_{x}^{\textrm{PMMH}}$ state particles, and the new invariant path is drawn from this set. If $r$ is small, however, the set of possible trajectories will be small, especially if there is path degeneracy in the particle filter. To counteract this issue, we first switch from a PMMH kernel to a PG kernel with $N_{x}^{\textrm{PMMH}}$ state particles, and then switch to a PG kernel with $N_{x}^{\textrm{PG}}$ state particles. The process used to switch between two different PG kernels is outlined in Section \ref{sec:switching_mechanics} and justified in Appendix \ref{app:kswitch}.
		
	All results are calculated based on $50$ independent runs. Computation is stopped if the runtime for any of the $50$ runs exceeded 2 weeks (336 hours). As a baseline, we run SMC$^2$ without switching for each model, targeting \eqref{eqn:da_pmmh_target}--\eqref{eqn:dt_pg_target}. Note that for density tempering, switching is done between \eqref{eqn:dt_pg_target} and \eqref{eqn:dt_pmmh_target_v2}, not \eqref{eqn:dt_pmmh_target} and \eqref{eqn:dt_pg_target}, as discussed in Section \ref{sec:dt_targets}. Due to the complications caused by \eqref{eqn:dt_pmmh_target_v2} for the reweighting step, \eqref{eqn:dt_pg_target} is always used as the default kernel in density tempering. 
		
	\subsection{Performance Metrics} \label{sec:scores}
	To evaluate the performance of our methods, we compare the mean squared error (MSE), averaged over the parameters, of the posterior mean from SMC$^2$ relative to the posterior mean from a particle MCMC run of length $1$ million. The number of state particles used for the latter is always $N_{x}^{\textrm{PMMH}}$. We use the total cost from the particle filter runs (PFC) as a measure of the computation cost. Each time the particle filter is run, PFC is incremented by $\Nx\times t$, where $\Nx$ is the number of state particles and $t$ is the number of observations. 
	
	We denote the MSE and PFC of the $j$th independent run of a method as $\textrm{MSE}_{\textrm{method}}^{j}$ and $\textrm{PFC}^{j}_{\textrm{method}}$, for $j = 1, \ldots, 50$. A penalty for each method is given as 
	\begin{align*}
		\Eff{method} = \frac{1}{50}\sum_{j=1}^{50}{{\textrm{PFC}}_{\textrm{method}}^{j}\cdot\textrm{MSE}_{\textrm{method}}^{j}},
	\end{align*}
	and the overall score relative to a base method is 
	\begin{align*}
		\RelEff{method} = \frac{\Eff{base}}{\Eff{method}}.
	\end{align*}
	Higher values are preferred and mean that the given method outperforms the base method in terms of accuracy and computation time. We also give a penalty for the MSE and PFC separately as 
	\begin{align*}
		\Eff{method, MSE} = \frac{1}{50}\sum_{j=1}^{50}{\textrm{MSE}_{\textrm{method}}^{j}}, \quad \Eff{method, PFC} = \frac{1}{50}\sum_{j=1}^{50}{{\textrm{PFC}}_{\textrm{method}}^{j}},
	\end{align*}
	with the relative scores given by 
	\begin{align*}
		\RelEff{method, MSE} = \frac{\Eff{base, MSE}}{\Eff{method, MSE}}, \quad \RelEff{method, PFC} = \frac{\Eff{base, PFC}}{\Eff{method, PFC}}.
	\end{align*}
	Again, higher values are preferred.
	
	\subsection{Brownian Motion Model} \label{sec:bm}
	Our first example is based on the stochastic differential equation
	\begin{align*}
		dX_t = \left(\beta - \frac{\gamma^2}{2}\right)dt + \gamma dB_t,
	\end{align*}
	where $B_t$ is a standard Brownian motion process. The observation and transition densities are
	\begin{align*}		
		g(y_t\mid x_{t}, \params) &= \mathcal{N}(x_{t}, \sigma^2), \\
		f(x_t\mid x_{t-1}, \params) &= \mathcal{N}\left(x_{t-1} + \beta - \frac{\gamma^2}{2}, \gamma^2\right), 
	\end{align*} 
	for $t = 1, \ldots, T$, with priors $\mathcal{N}(x_0\mid 3, 5^2)$, $\mathcal{N}(\beta\mid 2, 5^2)$, $\operatorname{Half-Normal}(\gamma\mid 2^2)$, and $\operatorname{Half-Normal}(\sigma\mid 2^2)$. We simulate $T=100$ observations from this model using $\params := (x_0, \beta, \gamma, \sigma) = (1, 1.2, 1.5, 1)$. The number of state particles used for the PMMH and PG kernels are $N_x^{\textrm{PMMH}} = 200$ and $N_x^{\textrm{PG}} = r\cdot200$ respectively, where the values for the latter are $200$ ($r = 1$), $100$ ($r = 0.5$), $40$ ($r = 0.2$) and $10$ ($r = 0.05$).
	
	Table \ref{tab:bm} shows the results for all the experiments. We find that runs with $\mathcal{K}_{\textrm{def}} = \textrm{PMMH}$ always have fewer intermediate targets than $\mathcal{K}_{\textrm{def}} = \textrm{PG}$, which is likely due to the marginalisation over the latent states in the reweighting step. It is interesting to note that the number of particle MCMC repeats ($R$) increases significantly as $N_x^{\textrm{PG}}$ decreases for density tempering SMC$^2$, but not for data annealing SMC$^2$. The high values of $R$ occur mostly in the initial stages of the algorithm and is likely due to the form of the target distribution when $g_d$ is small --- the initial targets in the sequence \eqref{eqn:dt_pg_target} are dominated by the density over the latent states. We observe this in all of the numerical examples.
	
	The best results for data annealing SMC$^2$ are when $\mathcal{K}_{\textrm{def}} = \textrm{PMMH}$. In this case, we are able to improve on the best performing of the standard methods, i.e.\ SMC$^2$ with a fixed PG kernel when $r=0.05$, by a factor of $1.9$. For density tempering SMC$^2$, $\mathcal{K}_{\textrm{def}} = \textrm{PMMH}$ with no switching has the best results, followed by $\mathcal{K}_{\textrm{def}} = \textrm{PG}$ with $r = 0.05$. This is unsurprising as the average number of intermediate targets\slash distributions is 7 for $\mathcal{K}_{\textrm{def}} = \textrm{PMMH}$ and 20 for $\mathcal{K}_{\textrm{def}} = \textrm{PG}$. There is little difference between always and lag-based switching when $\mathcal{K}_{\textrm{def}} = \textrm{PMMH}$, but the latter can significantly improve the results when $\mathcal{K}_{\textrm{def}} = \textrm{PG}$.

	\begin{table}[htp]
		\centering
		\small
		\begin{tabular}{|cccc|cc|ccc|}
			\hline
			Method & Test $\mathcal{K}_{\textrm{alt}}$ & $\mathcal{K}_{\textrm{def}}$ & $r$ & \# targets & $R$ (mean) & $\RelEff{MSE}$ & $\RelEff{PFC}$ & $\RelEff{}$ \\
			\hline
			DA & never & PMMH & 1.00 & 16 & 91 & 1.00 & 1.00 & 1.00 \\ 
			DA & never & PG & 1.00 & 91 & 56 & 0.69 & 0.24 & 0.15 \\ 
			DA & never & PG & 0.50 & 90 & 57 & 0.51 & 0.49 & 0.22 \\ 
			DA & never & PG & 0.20 & 90 & 62 & 0.86 & 1.19 & 0.85 \\ 
			DA & never & PG & 0.05 & 89 & 61 & 1.04 & 4.67 & 4.31 \\ 
			\hline 
			DA & always & PMMH & 1.00 & 15 & 59 & 1.05 & 1.98 & 1.86 \\ 
			DA & always & PMMH & 0.50 & 15 & 67 & 1.32 & 2.83 & 3.58 \\ 
			DA & always & PMMH & 0.20 & 15 & 78 & 0.96 & 5.27 & 4.70 \\ 
			DA & always & PMMH & 0.05 & 15 & 99 & 0.99 & 9.56 & 8.37 \\ 
			\hline 
			DA & always & PG & 1.00 & 89 & 40 & 0.87 & 0.32 & 0.24 \\ 
			DA & always & PG & 0.50 & 90 & 47 & 0.99 & 0.45 & 0.39 \\ 
			DA & always & PG & 0.20 & 89 & 50 & 0.57 & 0.78 & 0.40 \\ 
			DA & always & PG & 0.05 & 88 & 54 & 0.68 & 1.26 & 0.76 \\ 
			\hline 
			DA & lag & PMMH & 1.00 & 15 & 66 & 0.95 & 1.57 & 1.32 \\ 
			DA & lag & PMMH & 0.50 & 15 & 68 & 1.40 & 2.59 & 3.39 \\ 
			DA & lag & PMMH & 0.20 & 15 & 78 & 0.96 & 5.27 & 4.70 \\ 
			DA & lag & PMMH & 0.05 & 15 & 99 & 0.97 & 9.55 & 8.24 \\ 
			\hline 
			DA & lag & PG & 1.00 & 91 & 54 & 0.72 & 0.24 & 0.16 \\ 
			DA & lag & PG & 0.50 & 91 & 56 & 0.59 & 0.47 & 0.24 \\ 
			DA & lag & PG & 0.20 & 89 & 55 & 0.79 & 1.24 & 0.86 \\ 
			DA & lag & PG & 0.05 & 89 & 57 & 0.71 & 4.86 & 3.05 \\ 
			\hline 
			\hline 
			DT & never & PMMH & 1.00 & 7 & 36 & 0.86 & 1.84 & 1.41 \\ 
			DT & never & PG & 1.00 & 20 & 289 & 0.72 & 0.08 & 0.05 \\ 
			DT & never & PG & 0.50 & 20 & 289 & 0.68 & 0.16 & 0.10 \\ 
			DT & never & PG & 0.20 & 20 & 286 & 0.86 & 0.40 & 0.32 \\ 
			DT & never & PG & 0.05 & 20 & 302 & 0.85 & 1.52 & 1.23 \\ 
			\hline 
			DT & always & PG & 1.00 & 20 & 77 & 0.95 & 0.29 & 0.24 \\ 
			DT & always & PG & 0.50 & 20 & 79 & 0.79 & 0.31 & 0.22 \\ 
			DT & always & PG & 0.20 & 20 & 175 & 0.92 & 0.38 & 0.31 \\ 
			DT & always & PG & 0.05 & 21 & 282 & 0.68 & 1.04 & 0.63 \\ 
			\hline 
			DT & lag & PG & 1.00 & 20 & 77 & 0.76 & 0.29 & 0.20 \\ 
			DT & lag & PG & 0.50 & 20 & 79 & 0.78 & 0.32 & 0.22 \\ 
			DT & lag & PG & 0.20 & 20 & 185 & 0.78 & 0.41 & 0.27 \\ 
			DT & lag & PG & 0.05 & 20 & 290 & 0.61 & 1.31 & 0.72 \\ 
			\hline 
		\end{tabular}
		\caption{Results for the Brownian motion model applied to synthetic data. DA refers to data annealing and DT to density tempering. The column `$r$' is the fraction of state particles used for the PG kernel, where $N_{x}^{\textrm{PG}} = r\cdot N_{x}^{\textrm{PMMH}}$. For this model $N_{x}^{\textrm{PMMH}} = 200$. The column `\# targets' gives the average number of SMC$^2$ targets\slash distributions over the 50 independent runs. The column `$R$' refers to the average number of particle MCMC kernel applications over the 50 runs --- these are averaged over the number of targets\slash distributions also. The remaining 3 columns give the scores for the accuracy, computation cost and overall efficiency as defined in Section \ref{sec:scores}. }
		\label{tab:bm}
	\end{table}
	
	\subsection{Flexible-Allee Logistic Model} \label{sec:fa}
	Our second example is the flexible-allee logistic model of \citet{Peters2010} 
	\begin{align*}		
		g(y_t\mid x_{t}, \params) &= \mathcal{N}(y_t\mid \log{(x_t)}, \sigma^2), \\
		\log{(x_{t+1})} &=\log{(x_{t})} + \beta_0 + \beta_1 x_{t} + \beta_2 x_{t}^2 + z_t, \quad z_t \sim\mathcal{N}(0, \gamma^2),
	\end{align*}
	which we apply to data measuring monthly female nutria abundance in East Anglia. Unlike a logistic growth model, which assumes a higher growth rate for a smaller population (e.g.\ due to less competition for resources), the allee effect assumes the opposite. Due to computational constraints, the model is fit to the first $60$ observations of the dataset. The priors assigned are $\beta_0 \sim \mathcal{N}(0, 0.2^2)$, $\beta_1, \beta_2 \sim \mathcal{N}(0, 0.001^2)$, $\operatorname{Half-Normal}(x_0\mid 1000^2)$ and $\gamma, \sigma \sim \operatorname{Exp}(1)$. The number of state particles used for the PMMH and PG kernels are $N_{x}^{\textrm{PMMH}} = 1700$ and $N_{x}^{\textrm{PG}} \in \{1700, 850, 340, 85\}$, the latter corresponding to $r = 1, 0.5, 0.2, 0.05$. 
	
	Results for all experiments are shown in Table \ref{tab:fa}. Again, runs with $\mathcal{K}_{\textrm{def}} = \textrm{PMMH}$ have less intermediate targets. As before, data annealing SMC$^2$ with $\mathcal{K}_{\textrm{def}} = \textrm{PMMH}$ gives the best results in terms of overall efficiency. On this example, we are able to improve on the best performing of the standard methods by a factor of $2$ for both data annealing and density tempering, despite the higher number of intermediate targets\slash distributions for the latter. Lag-based switching has poorer performance for this model compared to always switching, although it can still outperform the standard methods.
	
	\begin{table}[htp]
		\centering
		\small
		\begin{tabular}{|cccc|cc|ccc|}
			\hline
			Method & Test $\mathcal{K}_{\textrm{alt}}$ & $\mathcal{K}_{\textrm{def}}$ & $r$ & \# targets & $R$ (mean) & $\RelEff{MSE}$ & $\RelEff{PFC}$ & $\RelEff{}$ \\
			\hline
			DA & never & PMMH & 1.00 & 21 & 346 & 1.00 & 1.00 & 1.00 \\ 
			DA & never & PG & 1.00 & 60 & 469 & 1.49 & 0.15 & 0.23 \\ 
			DA & never & PG & 0.50 & 60 & 471 & 1.32 & 0.31 & 0.41 \\ 
			DA & never & PG & 0.20 & 60 & 489 & 1.37 & 0.78 & 1.05 \\ 
			DA & never & PG & 0.05 & 60 & 487 & 1.14 & 3.09 & 3.52 \\ 
			\hline 
			DA & always & PMMH & 1.00 & 21 & 383 & 1.36 & 0.99 & 1.37 \\ 
			DA & always & PMMH & 0.50 & 21 & 353 & 1.35 & 1.27 & 1.68 \\ 
			DA & always & PMMH & 0.20 & 21 & 627 & 1.34 & 1.94 & 2.62 \\ 
			DA & always & PMMH & 0.05 & 21 & 815 & 1.29 & 5.77 & 7.15 \\ 
			\hline 
			DA & always & PG & 1.00 & 60 & 149 & 1.15 & 0.55 & 0.61 \\ 
			DA & always & PG & 0.50 & 60 & 177 & 1.16 & 0.65 & 0.75 \\ 
			DA & always & PG & 0.20 & 60 & 324 & 1.10 & 0.94 & 0.99 \\ 
			DA & always & PG & 0.05 & 60 & 431 & 1.42 & 2.15 & 2.97 \\ 
			\hline 
			DA & lag & PMMH & 1.00 & 21 & 394 & 1.02 & 0.95 & 1.00 \\ 
			DA & lag & PMMH & 0.50 & 21 & 355 & 1.34 & 1.07 & 1.43 \\ 
			DA & lag & PMMH & 0.20 & 21 & 625 & 0.98 & 1.21 & 1.14 \\ 
			DA & lag & PMMH & 0.05 & 21 & 848 & 1.42 & 3.58 & 4.71 \\ 
			\hline 
			DA & lag & PG & 1.00 & 60 & 229 & 1.58 & 0.19 & 0.30 \\ 
			DA & lag & PG & 0.50 & 60 & 297 & 2.01 & 0.31 & 0.63 \\ 
			DA & lag & PG & 0.20 & 60 & 412 & 1.27 & 0.73 & 0.87 \\ 
			DA & lag & PG & 0.05 & 60 & 473 & 1.76 & 3.10 & 5.35 \\ 
			\hline 
			\hline 
			DT & never & PMMH & 1.00 & 13 & 875 & 1.16 & 0.10 & 0.12 \\ 
			DT & never & PG & 1.00 & - & - & - & - & - \\ 
			DT & never & PG & 0.50 & - & - & - & - & - \\ 
			DT & never & PG & 0.20 & - & - & - & - & - \\ 
			DT & never & PG & 0.05 & - & - & - & - & - \\ 
			\hline 
			DT & always & PG & 1.00 & 24 & 299 & 1.28 & 0.14 & 0.18 \\ 
			DT & always & PG & 0.50 & 24 & 347 & 1.48 & 0.13 & 0.21 \\ 
			DT & always & PG & 0.20 & 24 & 398 & 1.42 & 0.17 & 0.24 \\ 
			DT & always & PG & 0.05 & 24 & 1750 & 1.74 & 0.20 & 0.28 \\ 
			\hline 
			DT & lag & PG & 1.00 & 24 & 300 & 1.46 & 0.14 & 0.21 \\ 
			DT & lag & PG & 0.50 & - & - & - & - & - \\ 
			DT & lag & PG & 0.20 & - & - & - & - & - \\ 
			DT & lag & PG & 0.05 & - & - & - & - & - \\  
			\hline 
		\end{tabular}
		\caption{Results for the flexible-allee logistic model applied to the nutria abundance dataset. DA refers to data annealing and DT to density tempering. The column `$r$' is the fraction of state particles used for the PG kernel, where $N_{x}^{\textrm{PG}} = r\cdot N_{x}^{\textrm{PMMH}}$. For this model $N_{x}^{\textrm{PMMH}} = 200$. The column `\# targets' gives the average number of SMC$^2$ targets\slash distributions over the 50 independent runs. The column `$R$' refers to the average number of particle MCMC kernel applications over the 50 runs --- these are averaged over the number of targets\slash distributions also. The remaining 3 columns give the scores for the accuracy, computation cost and overall efficiency as defined in Section \ref{sec:scores}. Dashed rows indicate that computation for at least one of the 50 independent runs exceeded 336 hours.}
		\label{tab:fa}
	\end{table}

	\subsection{Stochastic Volatility in Mean Model} \label{sec:svm}
	The third model considered is the stochastic volatility in mean model \citep{Koopman2002} given by
	\begin{align*}
		g(y_{t}\mid h_t, y_{t-1}, \params) = \mathcal{N}(a + b y_{t-1} + ds^2\exp{(h_t)}, s^2\exp{(h_t)}), \\
		f(h_{t}\mid h_{t-1}, \params) = \mathcal{N}(\phi h_{t-1}, \sigma^2), \quad h_1\sim \mathcal{N}\left(0, \frac{\sigma^2}{1 - \phi^2}\right).
	\end{align*}
	
	The priors used are $a, d\sim\mathcal{N}(0, 10^2)$, $b, \phi\sim\mathcal{U}(0, 1)$ and $\sigma, s\sim\operatorname{Half-Normal}(2^2)$. The model is fit to weekly and daily demeaned returns of the S\&P/TSX Composite index for the periods 07/07/2010-07/07/2015 and 07/07/2010-03/01/2013 respectively. The weekly and daily closing prices during these periods were downloaded from finance.yahoo \citep{finance.yahoo}, and the demeaned returns are calculated as 
	\begin{align*}
		y_t = 100\left(R_{t+1} - \frac{1}{T-1}\sum_{i=1}^{T-1}{R_{i+1}} \right), \quad R_{t+1} = \log{P_{t+1}} - \log{P_{t}}, \quad t = 1, 2, \ldots, T-1,
	\end{align*}
	where $P_t$ is the closing price at week or day $t$. The weekly returns dataset has length $T-1 = 260$ and the daily returns dataset has length $T-1 = 626$. The number of state particles used for the PMMH and PG kernels are $N_x^{\textrm{PMMH}} = 60$ and $N_x^{\textrm{PG}} \in \{60, 30, 12, 3\}$, i.e.\ $r = 1, 0.5, 0.2, 0.05$, for the weekly returns dataset and $N_x^{\textrm{PMMH}} = 200$ and $N_x^{\textrm{PG}} \in \{200, 100, 40, 10\}$, i.e.\ $r = 1, 0.5, 0.2, 0.05$, for the daily returns dataset.
	
	Tables \ref{tab:svmw} and \ref{tab:svmd} show the results for the weekly and daily returns respectively. As before, runs with $\mathcal{K}_{\textrm{def}} = \textrm{PMMH}$ have fewer intermediate targets than $\mathcal{K}_{\textrm{def}} = \textrm{PG}$. The runs with $\mathcal{K}_{\textrm{def}} = \textrm{PG}$ have fairly good results for the weekly returns data, but not for the daily returns. The poorer performance for the latter is likely due to path degeneracy as a result of the longer time series. In both cases, data annealing SMC$^2$ has the best results and is able to improve on the best performing of the standard methods by a factor of at least $1.5$ for the weekly returns, and by a factor of at least $4.3$ for the daily returns. For density tempering SMC$^2$, allowing the particle MCMC kernel to switch gives some improvement for the weekly returns, but no improvement for the daily returns. As in Section \ref{sec:bm}, there is little difference between always and lag-based switching when $\mathcal{K}_{\textrm{def}} = \textrm{PMMH}$, but the latter can improve the results when $\mathcal{K}_{\textrm{def}} = \textrm{PG}$.
	
	\begin{table}[htp]
		\centering
		\small
		\begin{tabular}{|cccc|cc|ccc|}
			\hline
			Method & Test $\mathcal{K}_{\textrm{alt}}$ & $\mathcal{K}_{\textrm{def}}$ & $r$ & \# targets & $R$ (mean) & $\RelEff{MSE}$ & $\RelEff{PFC}$ & $\RelEff{}$ \\
			\hline
			DA & never & PMMH & 1.00 & 25 & 717 & 1.00 & 1.00 & 1.00 \\ 
			DA & never & PG & 1.00 & 72 & 285 & 1.36 & 1.08 & 1.36 \\ 
			DA & never & PG & 0.50 & 72 & 286 & 1.27 & 2.14 & 2.59 \\ 
			DA & never & PG & 0.20 & 72 & 293 & 1.12 & 5.19 & 5.47 \\ 
			DA & never & PG & 0.05 & 72 & 328 & 0.82 & 17.59 & 13.62 \\ 
			\hline 
			DA & always & PMMH & 1.00 & 26 & 421 & 1.06 & 1.81 & 1.82 \\ 
			DA & always & PMMH & 0.50 & 26 & 454 & 0.95 & 3.15 & 2.92 \\ 
			DA & always & PMMH & 0.20 & 26 & 505 & 1.05 & 7.42 & 7.05 \\ 
			DA & always & PMMH & 0.05 & 26 & 592 & 1.07 & 20.27 & 20.65 \\ 
			\hline 
			DA & always & PG & 1.00 & 72 & 217 & 1.08 & 1.33 & 1.36 \\ 
			DA & always & PG & 0.50 & 72 & 249 & 0.93 & 2.03 & 1.70 \\ 
			DA & always & PG & 0.20 & 72 & 286 & 0.97 & 4.31 & 3.99 \\ 
			DA & always & PG & 0.05 & 72 & 329 & 0.83 & 10.48 & 8.20 \\ 
			\hline 
			DA & lag & PMMH & 1.00 & 26 & 698 & 1.41 & 1.04 & 1.43 \\ 
			DA & lag & PMMH & 0.50 & 26 & 538 & 1.07 & 1.96 & 2.22 \\ 
			DA & lag & PMMH & 0.20 & 26 & 487 & 0.94 & 7.66 & 6.65 \\ 
			DA & lag & PMMH & 0.05 & 26 & 604 & 0.99 & 20.13 & 18.99 \\ 
			\hline 
			DA & lag & PG & 1.00 & 72 & 228 & 0.87 & 1.20 & 1.02 \\ 
			DA & lag & PG & 0.50 & 72 & 259 & 1.12 & 2.17 & 2.27 \\ 
			DA & lag & PG & 0.20 & 72 & 298 & 1.34 & 4.84 & 6.06 \\ 
			DA & lag & PG & 0.05 & 72 & 334 & 1.00 & 16.97 & 15.85 \\ 
			\hline 
			\hline 
			DT & never & PMMH & 1.00 & 12 & 151 & 1.32 & 2.35 & 2.98 \\ 
			DT & never & PG & 1.00 & 16 & 1133 & 1.44 & 0.23 & 0.31 \\ 
			DT & never & PG & 0.50 & 16 & 1150 & 1.05 & 0.44 & 0.44 \\ 
			DT & never & PG & 0.20 & 16 & 1178 & 1.03 & 1.09 & 1.07 \\ 
			DT & never & PG & 0.05 & 16 & 1453 & 1.06 & 3.51 & 3.50 \\ 
			\hline 
			DT & always & PG & 1.00 & 16 & 125 & 0.93 & 2.01 & 1.77 \\ 
			DT & always & PG & 0.50 & 16 & 136 & 1.29 & 2.01 & 2.44 \\ 
			DT & always & PG & 0.20 & 16 & 172 & 1.47 & 2.26 & 3.16 \\ 
			DT & always & PG & 0.05 & 16 & 1142 & 0.98 & 3.58 & 3.30 \\ 
			\hline 
			DT & lag & PG & 1.00 & 16 & 126 & 1.01 & 1.98 & 1.86 \\ 
			DT & lag & PG & 0.50 & 16 & 137 & 1.41 & 2.01 & 2.66 \\ 
			DT & lag & PG & 0.20 & 16 & 174 & 1.08 & 2.29 & 2.34 \\ 
			DT & lag & PG & 0.05 & 16 & 1162 & 0.93 & 3.70 & 3.30 \\
			\hline 
		\end{tabular}
		\caption{Results for the stochastic volatility in mean model applied to the weekly returns dataset. DA refers to data annealing and DT to density tempering. The column `$r$' is the fraction of state particles used for the PG kernel, where $N_{x}^{\textrm{PG}} = r\cdot N_{x}^{\textrm{PMMH}}$. For this model $N_{x}^{\textrm{PMMH}} = 200$. The column `\# targets' gives the average number of SMC$^2$ targets\slash distributions over the 50 independent runs. The column `$R$' refers to the average number of particle MCMC kernel applications over the 50 runs --- these are averaged over the number of targets\slash distributions also. The remaining 3 columns give the scores for the accuracy, computation cost and overall efficiency as defined in Section \ref{sec:scores}.}
		\label{tab:svmw}
	\end{table}

	\begin{table}[htp]
		\centering
		\small
		\begin{tabular}{|cccc|cc|ccc|}
			\hline
			Method & Test $\mathcal{K}_{\textrm{alt}}$ & $\mathcal{K}_{\textrm{def}}$ & $r$ & \# targets & $R$ (mean) & $\RelEff{MSE}$ & $\RelEff{PFC}$ & $\RelEff{}$ \\
			\hline
			DA & never & PMMH & 1.00 & 26 & 218 & 1.00 & 1.00 & 1.00 \\ 
			DA & never & PG & 1.00 & 86 & 472 & 0.99 & 0.12 & 0.11 \\ 
			DA & never & PG & 0.50 & 86 & 477 & 0.80 & 0.23 & 0.18 \\ 
			DA & never & PG & 0.20 & 86 & 482 & 1.06 & 0.57 & 0.59 \\ 
			DA & never & PG & 0.05 & 86 & 526 & 0.77 & 2.06 & 1.55 \\ 
			\hline 
			DA & always & PMMH & 1.00 & 26 & 207 & 0.86 & 1.01 & 0.82 \\ 
			DA & always & PMMH & 0.50 & 26 & 251 & 1.06 & 1.20 & 1.24 \\ 
			DA & always & PMMH & 0.20 & 26 & 370 & 0.71 & 2.33 & 1.54 \\ 
			DA & always & PMMH & 0.05 & 26 & 469 & 1.02 & 6.59 & 6.58 \\ 
			\hline 
			DA & always & PG & 1.00 & 85 & 148 & 0.97 & 0.43 & 0.40 \\ 
			DA & always & PG & 0.50 & 86 & 168 & 0.99 & 0.45 & 0.44 \\ 
			DA & always & PG & 0.20 & 85 & 292 & 0.75 & 0.62 & 0.46 \\ 
			DA & always & PG & 0.05 & 87 & 510 & 0.47 & 1.56 & 0.72 \\ 
			\hline 
			DA & lag & PMMH & 1.00 & 26 & 230 & 0.95 & 0.91 & 0.86 \\ 
			DA & lag & PMMH & 0.50 & 26 & 235 & 1.24 & 1.02 & 1.25 \\ 
			DA & lag & PMMH & 0.20 & 26 & 357 & 1.21 & 1.70 & 2.06 \\ 
			DA & lag & PMMH & 0.05 & 26 & 473 & 1.07 & 6.51 & 6.64 \\ 
			\hline 
			DA & lag & PG & 1.00 & 86 & 157 & 0.97 & 0.40 & 0.39 \\ 
			DA & lag & PG & 0.50 & 87 & 465 & 0.68 & 0.23 & 0.15 \\ 
			DA & lag & PG & 0.20 & 87 & 475 & 0.83 & 0.56 & 0.45 \\ 
			DA & lag & PG & 0.05 & 86 & 526 & 0.60 & 2.05 & 1.20 \\ 
			\hline 
			\hline 
			DT & never & PMMH & 1.00 & 13 & 129 & 1.25 & 0.82 & 0.99 \\ 
			DT & never & PG & 1.00 & - & - & - & - & - \\ 
			DT & never & PG & 0.50 & - & - & - & - & - \\ 
			DT & never & PG & 0.20 & - & - & - & - & - \\ 
			DT & never & PG & 0.05 & - & - & - & - & - \\ 
			\hline 
			DT & always & PG & 1.00 & 19 & 143 & 0.88 & 0.52 & 0.45 \\ 
			DT & always & PG & 0.50 & 19 & 143 & 0.71 & 0.52 & 0.37 \\ 
			DT & always & PG & 0.20 & 19 & 193 & 1.32 & 0.55 & 0.69 \\ 
			DT & always & PG & 0.05 & - & - & - & - & - \\ 
			\hline 
			DT & lag & PG & 1.00 & 19 & 143 & 0.88 & 0.52 & 0.45 \\ 
			DT & lag & PG & 0.50 & 19 & 143 & 0.71 & 0.52 & 0.37 \\ 
			DT & lag & PG & 0.20 & 19 & 198 & 1.12 & 0.55 & 0.59 \\ 
			DT & lag & PG & 0.05 & - & - & - & - & - \\ 
			\hline 
		\end{tabular}
		\caption{Results for the stochastic volatility in mean model applied to the daily returns dataset. DA refers to data annealing and DT to density tempering. The column `$r$' is the fraction of state particles used for the PG kernel, where $N_{x}^{\textrm{PG}} = r\cdot N_{x}^{\textrm{PMMH}}$. For this model $N_{x}^{\textrm{PMMH}} = 200$. The column `\# targets' gives the average number of SMC$^2$ targets\slash distributions over the 50 independent runs. The column `$R$' refers to the average number of particle MCMC kernel applications over the 50 runs --- these are averaged over the number of targets\slash distributions also. The remaining 3 columns give the scores for the accuracy, computation cost and overall efficiency as defined in Section \ref{sec:scores}. Dashed rows indicate that computation for at least one of the 50 independent runs exceeded 336 hours.}
		\label{tab:svmd}
	\end{table}

	\subsection{First Order Autoregressive Model} \label{sec:ar1}
	Our final model uses a stationary AR(1) process to model the latent states, and uses the observation density of the univariate Ornstein-Uhlenbeck example of \citet{Mendes2020},
	\begin{align*}
		y_t &\sim \mathcal{N}\left(z^{\top}_t\vect{\beta}, \exp{(x_t)}\right), \\
		x_t\mid x_{t-1} &\sim \mathcal{N}\left(\mu + \phi(x_{t-1} - \mu), \sigma^2\right), \\
		x_1 &\sim \mathcal{N}\left(\mu, \frac{\sigma^2}{1-\phi^2}\right).
	\end{align*}
	The parameters are $0 < \phi < 1$, $\sigma > 0$, $\mu$ and $\vect{\beta} \in \mathbb{R}^{d_{\beta} \times 1}$. The priors assigned are $\phi\sim\mathcal{U}(0, 1)$, $\sigma\sim\operatorname{Half-Normal}(10^2)$, $\mu\sim\mathcal{N}(0, 5^2)$ and $\beta \sim \mathcal{N}(\vect{0}, I_{d_{\beta}})$, where $\vect{0}$ is a vector of length $d_{\beta}$ containing all zeros, and $I_{d_{\beta}}$ is the identity matrix of size $I_{d_{\beta}}\times I_{d_{\beta}}$.

	We simulate $400$ observations with $d_{\beta} = 20$ using $\phi = 0.5$, $\sigma = 1$, $\mu = 0.38$ and $\vect{\beta} = (0.1, 0.1, \cdots, 0.1)^{\top}$. The covariates are simulated as $z_t \sim \mathcal{N}(0, I_{d_{\beta}})$. The number of state particles used for the PMMH and PG kernels are $N_x^{\textrm{PMMH}} = 420$ and $N_x^{\textrm{PG}} \in \{420, 210, 84, 21\}$, i.e.\ $r = 1, 0.5, 0.2, 0.05$. 
	
	Table \ref{tab:ar1} shows the results for this model. Due to the higher number of parameters in the observation model, PG is expected to outperform PMMH for this example. We observe similar results as for the previous examples. Allowing the mutation kernel to switch can improve the best performing of the standard methods by a factor of at least $6.4$ for data annealing and $1.6$ for density tempering SMC$^2$. As before, lag-based switching makes little difference when $\mathcal{K}_{\textrm{def}} = \textrm{PMMH}$, but can improve the results when $\mathcal{K}_{\textrm{def}} = \textrm{PG}$.

	\begin{table}[htp]
		\centering
		\small
		\begin{tabular}{|cccc|cc|ccc|}
			\hline
			Method & Test $\mathcal{K}_{\textrm{alt}}$ & $\mathcal{K}_{\textrm{def}}$ & $r$ & \# targets & $R$ (mean) & $\RelEff{MSE}$ & $\RelEff{PFC}$ & $\RelEff{}$ \\
			\hline
			DA & never & PMMH & 1.00 & 114 & 449 & 1.00 & 1.00 & 1.00 \\ 
			DA & never & PG & 1.00 & - & - & - & - & - \\ 
			DA & never & PG & 0.50 & - & - & - & - & - \\ 
			DA & never & PG & 0.20 & 199 & 304 & 1.18 & 2.77 & 3.28 \\ 
			DA & never & PG & 0.05 & - & - & - & - & - \\ 
			\hline 
			DA & always & PMMH & 1.00 & 117 & 239 & 1.02 & 1.64 & 1.68 \\ 
			DA & always & PMMH & 0.50 & 117 & 279 & 1.07 & 2.58 & 2.78 \\ 
			DA & always & PMMH & 0.20 & 116 & 316 & 0.92 & 5.68 & 5.22 \\ 
			DA & always & PMMH & 0.05 & 116 & 373 & 1.12 & 16.23 & 18.23 \\ 
			\hline 
			DA & always & PG & 1.00 & 200 & 199 & 1.04 & 0.75 & 0.78 \\ 
			DA & always & PG & 0.50 & 200 & 226 & 1.04 & 1.14 & 1.18 \\ 
			DA & always & PG & 0.20 & - & - & - & - & - \\ 
			DA & always & PG & 0.05 & 199 & 299 & 1.18 & 6.93 & 8.20 \\ 
			\hline 
			DA & lag & PMMH & 1.00 & 114 & 430 & 1.14 & 1.01 & 1.15 \\ 
			DA & lag & PMMH & 0.50 & 116 & 347 & 0.88 & 1.22 & 1.07 \\ 
			DA & lag & PMMH & 0.20 & 116 & 325 & 0.93 & 3.64 & 3.53 \\ 
			DA & lag & PMMH & 0.05 & 116 & 364 & 1.20 & 16.38 & 19.69 \\ 
			\hline 
			DA & lag & PG & 1.00 & - & - & - & - & - \\ 
			DA & lag & PG & 0.50 & - & - & - & - & - \\ 
			DA & lag & PG & 0.20 & 200 & 265 & 1.01 & 2.75 & 2.77 \\ 
			DA & lag & PG & 0.05 & 200 & 295 & 1.15 & 10.94 & 12.55 \\ 
			\hline 
			\hline 
			DT & never & PMMH & 1.00 & 19 & 255 & 0.96 & 2.16 & 2.07 \\ 
			DT & never & PG & 1.00 & - & - & - & - & - \\ 
			DT & never & PG & 0.50 & - & - & - & - & - \\ 
			DT & never & PG & 0.20 & - & - & - & - & - \\ 
			DT & never & PG & 0.05 & - & - & - & - & - \\ 
			\hline 
			DT & always & PG & 1.00 & 27 & 213 & 1.10 & 1.76 & 1.91 \\ 
			DT & always & PG & 0.50 & 27 & 218 & 1.02 & 1.85 & 1.88 \\ 
			DT & always & PG & 0.20 & 27 & 388 & 1.00 & 2.19 & 2.20 \\ 
			DT & always & PG & 0.05 & - & - & - & - & - \\ 
			\hline 
			DT & lag & PG & 1.00 & 27 & 213 & 1.05 & 1.75 & 1.84 \\ 
			DT & lag & PG & 0.50 & 27 & 233 & 1.09 & 1.82 & 1.98 \\ 
			DT & lag & PG & 0.20 & 27 & 518 & 1.07 & 1.98 & 2.13 \\ 
			DT & lag & PG & 0.05 & - & - & - & - & - \\ 
			\hline 
		\end{tabular}
		\caption{Results for the AR(1) model applied to synthetic data. DA refers to data annealing and DT to density tempering. The column `$r$' is the fraction of state particles used for the PG kernel, where $N_{x}^{\textrm{PG}} = r\cdot N_{x}^{\textrm{PMMH}}$. For this model $N_{x}^{\textrm{PMMH}} = 200$. The column `\# targets' gives the average number of SMC$^2$ targets\slash distributions over the 50 independent runs. The column `$R$' refers to the average number of particle MCMC kernel applications over the 50 runs --- these are averaged over the number of targets\slash distributions also. The remaining 3 columns give the scores for the accuracy, computation cost and overall efficiency as defined in Section \ref{sec:scores}. Dashed rows indicate that computation for at least one of the 50 independent runs exceeded 336 hours.}
		\label{tab:ar1}
	\end{table}

	\section{Discussion} \label{sec:discussion}
	We introduce a method to automatically select between a PMMH and PG kernel at each iteration of SMC$^2$. Recall that one of the kernels is the default kernel, and the other is the alternate kernel. In the mutation step, both kernels are tested and the most efficient one is used to mutate the particles. Once the mutation is finished, the kernel (and therefore the sequence of targets) is switched back to the default kernel. We introduce a variation where the alternate kernel is tested based on a lag parameter. We find that this variation either has similar or improved performance compared to testing the alternate kernel at every iteration. As a result, we recommend using lag-based switching in general.
	
	Since the conventional sequence of targets for density tempering SMC$^2$ has an unknown marginal distribution when the tempering parameter is not equal to 0 or 1, PG is always used as the default kernel when applying our methodology in this setting. However, SMC$^2$ with PG as the default kernel generally has a higher number of intermediate targets than SMC$^2$ with PMMH as the default kernel, since the reweighting step for the former depends on the latent state trajectories. Consequently, the results are not as promising for density tempering SMC$^2$ as they are for data annealing SMC$^2$. 
	
	Unsurprisingly, the best results for data annealing SMC$^2$ are obtained when the default kernel is PMMH. We find that switching between particle MCMC kernels can significantly improve the standard method, i.e.\ data annealing SMC$^2$ with a fixed PMMH or PG kernel. Notably, the results for the AR(1) model and the stochastic volatility in mean model applied to the daily returns data show that our method can give improvement even when one of the candidate kernels is expected to perform poorly in general, e.g.\ PMMH when $\theta$ is high dimensional, or PG when the time series is very long. 
	
	In general, we recommend using PMMH as the default kernel where possible, as we find that it greatly improves the variance of the parameter particle weights, leading to less intermediate targets overall. For density tempering SMC$^2$ our approach can improve the results when using a fixed PG kernel. If PG mutations are desired, or expected to outperform PMMH, then we recommend allowing the kernels to switch between PG and PMMH. If PMMH is expected to give good performance, then standard density tempering SMC$^2$ with a fixed PMMH kernel may be preferred. It is worth noting, however, that data annealing SMC$^2$ when the kernels are allowed to switch outperforms density tempering SMC$^2$ on all of our examples. 
	
	Other methods which combine PMMH and PG, albeit in the context of particle MCMC samplers, are developed by \citet{Mendes2020} and \citet{Gunawan2017}. These methods introduce hybrid PMMH and PG algorithms, where a subset of the parameters is updated using PMMH and a subset is updated using PG. Our approach differs in that we use either PMMH or PG for all of the parameters. An area of future work is to investigate how these hybrid algorithms can be included in the set of candidate particle MCMC kernels for SMC$^2$. It would also be interesting to incorporate correlated particle MCMC \citep{Deligiannidis2018} kernels, including the correlated hybrid algorithm of \citet{Gunawan2017}, into our approach.
	
	Another avenue for future work is to develop strategies for adapting the tempering parameter when using our modified version of the PMMH sequence of targets. Another area of future work is to incorporate adaptation methods for the number of state particles into our methodology. Since our proposed approach allows each particle MCMC kernel to use a different number of state particles, one possibility is to include one or more candidate kernels with different values of $\Nx$. It would also be interesting to investigate our proposed methodology with other types of mutation kernels. 
	
	\section{Acknowledgments} \label{sec:ack}
	Imke Botha was supported by an Australian Research Training Program Stipend and a QUT Centre for Data Science Top-Up Scholarship. Christopher Drovandi was supported by an Australian Research Council Discovery Project (DP200102101). We gratefully acknowledge the computational resources provided by QUT's High Performance Computing and Research Support Group (HPC).

	\bibliographystyle{apalike}
	\bibliography{refs}

\begin{thebibliography}{}

\bibitem[Andrieu et~al., 2010]{Andrieu2010}
Andrieu, C., Doucet, A., and Holenstein, R. (2010).
\newblock {Particle Markov chain Monte Carlo methods}.
\newblock {\em Journal of the Royal Statistical Society: Series B (Statistical
  Methodology)}, 72(3):269--342.

\bibitem[Andrieu and Roberts, 2009]{Andrieu2009}
Andrieu, C. and Roberts, G.~O. (2009).
\newblock {The pseudo-marginal approach for efficient Monte Carlo
  computations}.
\newblock {\em The Annals of Statistics}, 37(2):697--725.

\bibitem[Betancourt, 2017]{Betancourt2017}
Betancourt, M. (2017).
\newblock {A conceptual introduction to Hamiltonian Monte Carlo}.
\newblock {\em arXiv preprint arXiv:1701.02434}.

\bibitem[Bon et~al., 2021]{Bon2021}
Bon, J.~J., Lee, A., and Drovandi, C. (2021).
\newblock {Accelerating sequential Monte Carlo with surrogate likelihoods}.
\newblock {\em Statistics and Computing}, 31(5).

\bibitem[Botha et~al., 2023]{Botha2023}
Botha, I., Kohn, R., South, L., and Drovandi, C. (2023).
\newblock Automatically adapting the number of state particles in {SMC}$^2$.
\newblock {\em Statistics and Computing}, 33(4).

\bibitem[Carson et~al., 2017]{Carson2017}
Carson, J., Crucifix, M., Preston, S., and Wilkinson, R.~D. (2017).
\newblock Bayesian model selection for the glacial{\textendash}interglacial
  cycle.
\newblock {\em Journal of the Royal Statistical Society: Series C (Applied
  Statistics)}, 67(1):25--54.

\bibitem[Chopin, 2002]{Chopin2002}
Chopin, N. (2002).
\newblock A sequential particle filter method for static models.
\newblock {\em Biometrika}, 89(3):539--552.

\bibitem[Chopin et~al., 2012]{Chopin2012}
Chopin, N., Jacob, P.~E., and Papaspiliopoulos, O. (2012).
\newblock {SMC}2: an efficient algorithm for sequential analysis of state space
  models.
\newblock {\em Journal of the Royal Statistical Society: Series B (Statistical
  Methodology)}, 75(3):397--426.

\bibitem[Chopin and Papaspiliopoulos, 2020]{Chopin2020}
Chopin, N. and Papaspiliopoulos, O. (2020).
\newblock {\em {An introduction to sequential Monte Carlo}}.
\newblock Springer International Publishing.

\bibitem[Chopin et~al., 2015]{Chopin2015a}
Chopin, N., Ridgway, J., Gerber, M., and Papaspiliopoulos, O. (2015).
\newblock {Towards automatic calibration of the number of state particles
  within the SMC$^2$ algorithm}.
\newblock {\em arXiv preprint arXiv:1506.00570}.

\bibitem[Chopin and Singh, 2015]{Chopin2015}
Chopin, N. and Singh, S.~S. (2015).
\newblock {On particle Gibbs sampling}.
\newblock {\em Bernoulli}, 21(3).

\bibitem[{Del Moral} et~al., 2006]{DelMoral2006}
{Del Moral}, P., Doucet, A., and Jasra, A. (2006).
\newblock {Sequential Monte Carlo samplers}.
\newblock {\em Journal of the Royal Statistical Society: Series B (Statistical
  Methodology)}, 68(3):411--436.

\bibitem[Deligiannidis et~al., 2018]{Deligiannidis2018}
Deligiannidis, G., Doucet, A., and Pitt, M.~K. (2018).
\newblock The correlated pseudo-marginal method.
\newblock {\em Journal of the Royal Statistical Society Series B: Statistical
  Methodology}, 80(5):839--870.

\bibitem[Doucet et~al., 2015]{Doucet2015}
Doucet, A., Pitt, M.~K., Deligiannidis, G., and Kohn, R. (2015).
\newblock Efficient implementation of {Markov chain Monte Carlo} when using an
  unbiased likelihood estimator.
\newblock {\em Biometrika}, 102(2):295--313.

\bibitem[Duan and Fulop, 2014]{Duan2014}
Duan, J.-C. and Fulop, A. (2014).
\newblock {Density-tempered marginalized sequential Monte Carlo samplers}.
\newblock {\em Journal of Business {\&} Economic Statistics}, 33(2):192--202.

\bibitem[Fearnhead and Taylor, 2013]{Fearnhead2013}
Fearnhead, P. and Taylor, B.~M. (2013).
\newblock {An adaptive sequential Monte Carlo sampler}.
\newblock {\em Bayesian Analysis}, 8(2):411--438.

\bibitem[Fulop et~al., 2022]{Fulop2022}
Fulop, A., Heng, J., and Li, J. (2022).
\newblock Efficient likelihood-based estimation via annealing for dynamic
  structural macrofinance models.
\newblock {\em arXiv preprint arXiv:2201.01094}.

\bibitem[Girolami and Calderhead, 2011]{Girolami2011}
Girolami, M. and Calderhead, B. (2011).
\newblock {Riemann manifold Langevin and Hamiltonian Monte Carlo methods}.
\newblock {\em Journal of the Royal Statistical Society: Series B (Statistical
  Methodology)}, 73(2):123--214.

\bibitem[Golightly and Kypraios, 2017]{Golightly2017}
Golightly, A. and Kypraios, T. (2017).
\newblock Efficient {SMC}2 schemes for stochastic kinetic models.
\newblock {\em Statistics and Computing}, 28(6):1215--1230.

\bibitem[Gordon et~al., 1993]{Gordon1993}
Gordon, N.~J., Salmond, D.~J., and Smith, A. F.~M. (1993).
\newblock Novel approach to nonlinear/non-{G}aussian {B}ayesian state
  estimation.
\newblock {\em {IEE} Proceedings F Radar and Signal Processing}, 140(2):107.

\bibitem[Gunawan et~al., 2017]{Gunawan2017}
Gunawan, D., Carter, C., and Kohn, R. (2017).
\newblock {Efficient Bayesian inference for multivariate factor stochastic
  volatility models with leverage}.
\newblock {\em arXiv preprint arXiv:1706.03938}.

\bibitem[Higham and Lin, 2013]{Higham2013}
Higham, N.~J. and Lin, L. (2013).
\newblock An improved {S}chur--{P}ad{\'{e}} algorithm for fractional powers of
  a matrix and their {F}r{\'{e}}chet derivatives.
\newblock {\em {SIAM} Journal on Matrix Analysis and Applications},
  34(3):1341--1360.

\bibitem[Koopman and Uspensky, 2002]{Koopman2002}
Koopman, S.~J. and Uspensky, E.~H. (2002).
\newblock The stochastic volatility in mean model: empirical evidence from
  international stock markets.
\newblock {\em Journal of Applied Econometrics}, 17(6):667--689.

\bibitem[Lindsten et~al., 2014]{Lindsten2014}
Lindsten, F., Jordan, M.~I., and Sch{\"o}n, T.~B. (2014).
\newblock {Particle Gibbs with ancestor sampling}.
\newblock {\em Journal of Machine Learning Research}, 15:2145--2184.

\bibitem[Lindsten and Schön, 2012]{Lindsten2012}
Lindsten, F. and Schön, T.~B. (2012).
\newblock {On the use of backward simulation in the particle Gibbs sampler}.
\newblock In {\em 2012 {IEEE} International Conference on Acoustics, Speech and
  Signal Processing ({ICASSP})}, pages 3845--3848. {IEEE}.

\bibitem[Liu and Chen, 1998]{Liu1998}
Liu, J.~S. and Chen, R. (1998).
\newblock Sequential {M}onte {C}arlo methods for dynamic systems.
\newblock {\em Journal of the American Statistical Association},
  93(443):1032--1044.

\bibitem[Mendes et~al., 2020]{Mendes2020}
Mendes, E.~F., Carter, C.~K., Gunawan, D., and Kohn, R. (2020).
\newblock {A flexible particle Markov chain Monte Carlo method}.
\newblock {\em Statistics and Computing}, 30(4):783--798.

\bibitem[Pasarica and Gelman, 2010]{Pasarica2010}
Pasarica, C. and Gelman, A. (2010).
\newblock Adaptively scaling the {Metropolis} algorithm using expected squared
  jumped distance.
\newblock {\em Statistica Sinica}, 20(1):343--364.

\bibitem[Peters et~al., 2010]{Peters2010}
Peters, G.~W., Hosack, G.~R., and Hayes, K.~R. (2010).
\newblock {Ecological non-linear state space model selection via adaptive
  particle Markov chain Monte Carlo (AdPMCMC)}.
\newblock {\em arXiv preprint arXiv:1005.2238}.

\bibitem[Pitt et~al., 2012]{Pitt2012}
Pitt, M.~K., dos Santos~Silva, R., Giordani, P., and Kohn, R. (2012).
\newblock {On some properties of Markov chain Monte Carlo simulation methods
  based on the particle filter}.
\newblock {\em Journal of Econometrics}, 171(2):134--151.

\bibitem[Price, 1971]{Price1971}
Price, G.~R. (1971).
\newblock Extension of the {Hardy-Weinberg Law} to assortative mating.
\newblock {\em Annals of Human Genetics}, 34(4):455--458.

\bibitem[{Rimella} et~al., 2022]{Rimella2022}
{Rimella}, L., {Alderton}, S., {Sammarro}, M., {Rowlingson}, B., {Cocker}, D.,
  {Feasey}, N., {Fearnhead}, P., and {Jewell}, C. (2022).
\newblock {Inference on Extended-Spectrum Beta-Lactamase Escherichia coli and
  Klebsiella pneumoniae data through SMC$^2$}.
\newblock {\em arXiv preprint arXiv:2208.11331}.

\bibitem[Roberts and Rosenthal, 2001]{Roberts2001}
Roberts, G.~O. and Rosenthal, J.~S. (2001).
\newblock {Optimal scaling for various Metropolis-Hastings algorithms}.
\newblock {\em Statistical Science}, 16(4).

\bibitem[Salomone et~al., 2018]{Salomone2018}
Salomone, R., South, L.~F., Drovandi, C.~C., and Kroese, D.~P. (2018).
\newblock {Unbiased and consistent nested sampling via sequential Monte Carlo}.
\newblock {\em arxiv preprint arXiv:1805.03924}.

\bibitem[Sherlock et~al., 2015]{Sherlock2015}
Sherlock, C., Thiery, A.~H., Roberts, G.~O., and Rosenthal, J.~S. (2015).
\newblock On the efficiency of pseudo-marginal random walk {M}etropolis
  algorithms.
\newblock {\em The Annals of Statistics}, 43(1):238--275.

\bibitem[Svensson et~al., 2015]{Svensson2015}
Svensson, A., Schön, T.~B., and Kok, M. (2015).
\newblock Nonlinear state space smoothing using the conditional particle
  filter.
\newblock {\em {IFAC}-{PapersOnLine}}, 48(28):975--980.

\bibitem[Whiteley, 2010]{Whiteley2010}
Whiteley, N. (2010).
\newblock {Discussion on particle Markov chain Monte Carlo methods}.
\newblock {\em Journal of the Royal Statistical Society: Series B (Statistical
  Methodology)}, 72(3):306--307.

\bibitem[{Yahoo Finance}, 2023]{finance.yahoo}
{Yahoo Finance} (2023).
\newblock {S\&P/TSX Composite index ({\textasciicircum}GSPTSE)}.
\newblock \url{https://finance.yahoo.com/quote/^GSPTSE/history?p=^GSPTSE}.

\end{thebibliography}
	
	\appendix
	
	\section{Kernel Switching Maths} \label{app:kswitch}
	This section presents the theory underpinning our approach for switching between particle MCMC kernels in SMC$^2$. We use the generalised importance sampling method of \citet{DelMoral2006} to switch between the different target distributions. At iteration $d$, the aim is to replace the set of variables $a$ with the set $b$. To reweight the parameter particles from $\gamma_{\textrm{current}}(a, \cdot)$ to $\gamma_{\textrm{new}}(d, \cdot)$, the incremental weights are 
	\begin{align*}
		IW = \frac{\gamma_{d, \textrm{new}}(\params, b)}{\gamma_{d, \textrm{current}}( \params, a)}\frac{L(b, a)}{K(a, b)} = \frac{\gamma_{d, \textrm{new}}(\params, b)}{\gamma_{d, \textrm{current}}( \params, a)}\operatorname{KR}_d,
	\end{align*}
	where $L(b, a)$ is the backward kernel, $K(a, b)$ is the forward kernel and $\operatorname{KR}_d$ is the kernel ratio. Once the incremental weights are calculated, the new parameter particle weights are 
	\begin{align*}
		w_d^n = w_d^n \cdot IW, \quad W_d^n = \frac{w_d^n}{\sum_{i=1}^{\Nt}w_d^i}, \quad n = 1,\ldots, \Nt.
	\end{align*}
	Notably, if $\operatorname{KR}_d = \gamma_{d, \textrm{current}}( \params, a)\slash \gamma_{d, \textrm{new}}( \params, a)$, the incremental weights become $1$.
	
	The following sections show how to switch between different particle MCMC kernels for data annealing and density tempering SMC$^2$. Through careful choice of forward and backward kernels, we show that $\operatorname{KR}_d = \gamma_{d, \textrm{current}}( \params, a)\slash \gamma_{d, \textrm{new}}( \params, a)$ in all cases. 
	
	In particular, our approach relies on the following approximation to the optimal backward kernel (see Sections 3.3.1 and 3.3.2 of \citealt{DelMoral2006}), 
	\begin{align}
		L(b, a) = \frac{\gamma_{d, \textrm{current}}( \params, a)K(a, b)}{\int{\gamma_{d, \textrm{current}}( \params, a)K(a, b)}da}, \label{eqn:approx_opt_bk}
	\end{align}
	which gives the kernel ratio as
	\begin{align}
		\operatorname{KR}_d = \frac{L(b, a)}{K(a, b)} = \frac{\gamma_{d, \textrm{current}}( \params, a)}{\int{\gamma_{d, \textrm{current}}( \params, a)K(a, b)}da}. \label{eqn:kr}
	\end{align}
	When switching to PMMH, evaluating the integral in \eqref{eqn:approx_opt_bk} requires the exact likelihood. In this case, we replace the exact likelihood with a particle filter estimator.

	\subsection{Data Annealing}
	First, recall that the $d$th target in a data annealing sequence of targets when using PMMH is 
		\begin{equation}
		\begin{aligned}
			\gamma_d(\params, x_{1:d}^{1:N_x^{\textrm{PMMH}}}, y_{1:d}) \ &= \ p(\params)\widehat{p_{N_x^{\textrm{PMMH}}}}(y_{1:d}\mid\params, x_{1:d}^{1:N_x^{\textrm{PMMH}}})\psi(x_{1:d}^{1:N_x^{\textrm{PMMH}}}\mid \params, y_{1:d}) \\
			&= \ p(\params)\widehat{p_{N_x^{\textrm{PMMH}}}}(y_{1:d}, x_{1:d}^{1:N_x^{\textrm{PMMH}}}\mid\params),
		\end{aligned}
		\label{aeq:da_pmmh_target}
	\end{equation}
	and when using PG is
	\begin{equation}
		\begin{aligned}
			&\gamma_d(\params, {k_{1:d}}, x_{1:d}^{k_{1:d}}, x_{1:d}^{1:N_x^{\textrm{PG}}, -{k_{1:d}}}, y_{1:d}) \\ 
			&= \ p(\params)p(x_{1:d}^{k_{1:d}}\mid \params)p(y_{1:d}\mid\params, x_{1:d}^{k_{1:d}}) \psi_d(x_{1:d}^{1:N_x^{\textrm{PG}}, -{k_{1:d}}}\mid\params, y_{1:d}, x_{1:d}^{k_{1:d}}, {k_{1:d}}) \\
			&\qquad \times\psi_d({k_{1:d}}\mid\params, y_{1:d}, x_{1:d}^{k_{1:d}}, x_{1:d}^{1:N_x^{\textrm{PG}}, -{k_{1:d}}}) \\
			&= \ p(\params)p(x_{1:d}^{k_{1:d}}\mid \params)p(y_{1:d}\mid\params, x_{1:d}^{k_{1:d}}) \psi_d(x_{1:d}^{1:N_x^{\textrm{PG}}, -{k_{1:d}}}, {k_{1:d}}\mid\params, y_{1:d}, x_{1:d}^{k_{1:d}}).
		\end{aligned}
		\label{aeq:da_pg_target}
	\end{equation}
	The following four sections describe how to switch between a PMMH and PG kernel, and how to switch between two different PMMH or two different PG kernels (where the number of state particles differ between kernels).
	
	\subsubsection{PG \texorpdfstring{$\rightarrow$}{->} PMMH}
	To reweight from \eqref{aeq:da_pg_target} to \eqref{aeq:da_pmmh_target}, i.e.\ to switch from a PG kernel to a PMMH kernel, we first need to obtain a new set of latent state trajectories, 
	\begin{align*}
		x_{1:d}^{1:N_x^{\textrm{PMMH}}} \sim \psi(x_{1:d}^{1:N_x^{\textrm{PMMH}}}\mid \params, y_{1:d}).
	\end{align*}
	The denominator of the kernel ratio in \eqref{eqn:kr} is
	\begin{align*}
		&\int{\left[\gamma_d(\params, {k_{1:d}}, x_{1:d}^{k_{1:d}}, x_{1:d}^{1:N_x^{\textrm{PG}}, -{k_{1:d}}}, y_{1:d})\psi(x_{1:d}^{1:N_x^{\textrm{PMMH}}}\mid \params, y_{1:d})\right]}d\{{k_{1:d}}, x_{1:d}^{k_{1:d}}, x_{1:d}^{1:N_x^{\textrm{PG}}, -{k_{1:d}}}\} \\
		&=\psi(x_{1:d}^{1:N_x^{\textrm{PMMH}}}\mid \params, y_{1:d})\int{\left[\gamma_d(\params, {k_{1:d}}, x_{1:d}^{k_{1:d}}, x_{1:d}^{1:N_x^{\textrm{PG}}, -{k_{1:d}}}, y_{1:d})\right]}d\{{k_{1:d}}, x_{1:d}^{k_{1:d}}, x_{1:d}^{1:N_x^{\textrm{PG}}, -{k_{1:d}}}\} \\
		&= p(\params)p(y_{1:d}\mid\params)\psi(x_{1:d}^{1:N_x^{\textrm{PMMH}}}\mid \params, y_{1:d}).
	\end{align*}
	The kernel ratio is then
	\begin{align*}
		\operatorname{KR}_d &= \frac{\gamma_d(\params, {k_{1:d}}, x_{1:d}^{k_{1:d}}, x_{1:d}^{1:N_x^{\textrm{PG}}, -{k_{1:d}}}, y_{1:d})}{p(\params)p(y_{1:d}\mid\params)\psi(x_{1:d}^{1:N_x^{\textrm{PMMH}}}\mid \params, y_{1:d})}.
	\end{align*}
	Replacing the exact likelihood $p(y_{1:d}\mid\params)$ with $\widehat{p_{N_x^{\textrm{PMMH}}}}(y_{1:d}\mid\params, x_{1:d}^{1:N_x^{\textrm{PMMH}}})$, gives
	\begin{align*}
		\operatorname{KR}_d = \frac{\gamma_d(\params, {k_{1:d}}, x_{1:d}^{k_{1:d}}, x_{1:d}^{1:N_x^{\textrm{PG}}, -{k_{1:d}}}, y_{1:d})}{\gamma_d(\params, x_{1:d}^{1:N_x^{\textrm{PMMH}}}, y_{1:d})}.
	\end{align*}
	
	\subsubsection{PMMH \texorpdfstring{$\rightarrow$}{->} PG}
	To reweight from \eqref{aeq:da_pmmh_target} to \eqref{aeq:da_pg_target}, i.e.\ to switch from a PMMH kernel to a PG kernel, a new latent state trajectory must be sampled. Running a standard particle filter, then drawing a new latent state trajectory gives the following forward kernel 
	\begin{align*}
		({k_{1:d}}, x_{1:d}^{k_{1:d}}, x_{1:d}^{1:N_x^{\textrm{PG}}, -{k_{1:d}}}) \sim \psi(x_{1:d}^{1:N_x^{\textrm{PG}}}\mid \params, y_{1:d})\psi_d({k_{1:d}}\mid\params, y_{1:d}, x_{1:d}^{k_{1:d}}, x_{1:d}^{1:N_x^{\textrm{PG}}, -{k_{1:d}}}).
	\end{align*}
	Using this forward kernel, the denominator of the kernel ratio in \eqref{eqn:kr} is 
	\begin{align*}
		&\int{\left[\gamma_d(\params, x_{1:d}^{1:N_x^{\textrm{PMMH}}}, y_{1:d})\psi(x_{1:d}^{1:N_x^{\textrm{PG}}}\mid \params, y_{1:d})\psi_d({k_{1:d}}\mid\params, y_{1:d}, x_{1:d}^{k_{1:d}}, x_{1:d}^{1:N_x^{\textrm{PG}}, -{k_{1:d}}})\right]}d\{x_{1:d}^{1:N_x^{\textrm{PMMH}}}\} \\
		&= \gamma_d(\params, y_{1:d})\psi(x_{1:d}^{1:N_x^{\textrm{PG}}}\mid \params, y_{1:d})\psi_d({k_{1:d}}\mid\params, y_{1:d}, x_{1:d}^{k_{1:d}}, x_{1:d}^{1:N_x^{\textrm{PG}}, -{k_{1:d}}}) \\
		&= p(\params)p(y_{1:d}\mid\params)\psi(x_{1:d}^{1:N_x^{\textrm{PG}}}\mid \params, y_{1:d})\psi_d({k_{1:d}}\mid\params, y_{1:d}, x_{1:d}^{k_{1:d}}, x_{1:d}^{1:N_x^{\textrm{PG}}, -{k_{1:d}}}).
	\end{align*}
	By noting that 
	\begin{align*}
		\psi_d(x_{1:d}^{1:N_x^{\textrm{PG}}, -{k_{1:d}}}, x_{1:d}^{k_{1:d}}, {k_{1:d}}\mid\params, y_{1:d}) &= \psi(x_{1:d}^{1:N_x^{\textrm{PG}}}\mid \params, y_{1:d})\psi_d({k_{1:d}}\mid\params, y_{1:d}, x_{1:d}^{k_{1:d}}, x_{1:d}^{1:N_x^{\textrm{PG}}, -{k_{1:d}}}) \\
		&= \psi_d(x_{1:d}^{1:N_x^{\textrm{PG}}, -{k_{1:d}}}, {k_{1:d}}\mid\params, y_{1:d}, x_{1:d}^{k_{1:d}})p(x_{1:d}^{k_{1:d}}\mid \params, y_{1:d}),
	\end{align*}
	the kernel ratio can be written as
	\begin{align*}
		\operatorname{KR}_d &= \frac{\gamma_d(\params, x_{1:d}^{1:N_x^{\textrm{PMMH}}}, y_{1:d})}{p(\params)p(y_{1:d}\mid\params)\psi_d(x_{1:d}^{1:N_x^{\textrm{PG}}, -{k_{1:d}}}, {k_{1:d}}\mid\params, y_{1:d}, x_{1:d}^{k_{1:d}})p(x_{1:d}^{k_{1:d}}\mid \params, y_{1:d})} \\
		&= \frac{\gamma_d(\params, x_{1:d}^{1:N_x^{\textrm{PMMH}}}, y_{1:d})}{\gamma_d(\params, {k_{1:d}}, x_{1:d}^{k_{1:d}}, x_{1:d}^{1:N_x^{\textrm{PG}}, -{k_{1:d}}}, y_{1:d})}.
	\end{align*}
	
	\subsubsection{PMMH \texorpdfstring{$\rightarrow$}{->} PMMH}
	To switch between two different PMMH kernels, i.e.\ to switch from $\gamma_d(\params, x_{1:d}^{1:N_x^{\textrm{PMMH}, 1}}, y_{1:d})$ to $\gamma_d(\params, x_{1:d}^{1:N_x^{\textrm{PMMH}, 2}}, y_{1:d})$, we can use the method outlined in \citet{Botha2023}. Specifically, we draw a new set of latent state trajectories using a standard particle filter
	\begin{align*}
		x_{1:d}^{1:N_x^{\textrm{PMMH}, 2}} \sim \psi(x_{1:d}^{1:N_x^{\textrm{PMMH}, 2}}\mid \params, y_{1:d}),
	\end{align*}
	and use the backward kernel 
	\begin{align*}
		L = \frac{\widehat{p_{N_x^{\textrm{PMMH, 1}}}}(y_{1:d}\mid\params, x_{1:d}^{1:N_x^{\textrm{PMMH, 1}}})\psi(x_{1:d}^{1:N_x^{\textrm{PMMH}, 2}}\mid \params, y_{1:d})}{\widehat{p_{N_x^{\textrm{PMMH, 1}}}}(y_{1:d}\mid\params, x_{1:d}^{1:N_x^{\textrm{PMMH}, 2}})}.
	\end{align*}
	This choice of backward kernel gives incremental weights of $1$. See \citet{Botha2023} for more details.
	
	\subsubsection{PG \texorpdfstring{$\rightarrow$}{->} PG}
	To switch between two different PG kernels, i.e.\ to switch from $\gamma_d($$\params$$, $${k_{1:d}}$$, $$x_{1:d}^{k_{1:d}}$$, $$x_{1:d}^{1:N_x^{\textrm{PG}, 1}, -{k_{1:d}}}, $ $y_{1:d})$ to $\gamma_d(\params, {\tilde{k}_{1:d}}, \tilde{x}_{1:d}^{\tilde{k}_{1:d}}, \tilde{x}_{1:d}^{1:N_x^{\textrm{PG}, 2}, -{\tilde{k}_{1:d}}}, y_{1:d})$, we run the conditional particle filter and
	draw a new latent state trajectory
	\begin{align*}
		({\tilde{k}_{1:d}}, \tilde{x}_{1:d}^{\tilde{k}_{1:d}}, \tilde{x}_{1:d}^{1:N_x^{\textrm{PG}, 2}, -{\tilde{k}_{1:d}}}) &\sim \psi_d(\tilde{x}_{1:d}^{1:N_x^{\textrm{PG}, 2}, -{\tilde{k}_{1:d}}}, {\tilde{k}_{1:d}}\mid\params, y_{1:d}, x_{1:d}^{k_{1:d}}).
	\end{align*}
	The denominator of the kernel ratio in \eqref{eqn:kr} then becomes
	\begin{align*}
		&\int{\left[\gamma_d(\params, {k_{1:d}}, x_{1:d}^{k_{1:d}}, x_{1:d}^{1:N_x^{\textrm{PG}, 1}, -{k_{1:d}}}, y_{1:d})\right.} \\
		&\qquad\times{\left. \psi_d(\tilde{x}_{1:d}^{1:N_x^{\textrm{PG}, 2}, -{\tilde{k}_{1:d}}}, {\tilde{k}_{1:d}}\mid\params, y_{1:d}, x_{1:d}^{k_{1:d}})\right]}d\{{k_{1:d}}, x_{1:d}^{1:N_x^{\textrm{PG}, 1}, -{k_{1:d}}}\} \\
		&=\gamma_d(\params, x_{1:d}^{k_{1:d}}, {\tilde{k}_{1:d}}, \tilde{x}_{1:d}^{\tilde{k}_{1:d}}, \tilde{x}_{1:d}^{1:N_x^{\textrm{PG}, 2}, -{\tilde{k}_{1:d}}, -{k_{1:d}}}, y_{1:d}) \\
		&= \gamma_d(\params, {\tilde{k}_{1:d}}, \tilde{x}_{1:d}^{\tilde{k}_{1:d}}, \tilde{x}_{1:d}^{1:N_x^{\textrm{PG}, 2}, -{\tilde{k}_{1:d}}}, y_{1:d}).
	\end{align*}
	The kernel ratio is
	\begin{align*}
		\operatorname{KR}_d &= \frac{\gamma_d(\params, {k_{1:d}}, x_{1:d}^{k_{1:d}}, x_{1:d}^{1:N_x^{\textrm{PG}, 1}, -{k_{1:d}}}, y_{1:d})}{\gamma_d(\params, {\tilde{k}_{1:d}}, \tilde{x}_{1:d}^{\tilde{k}_{1:d}}, \tilde{x}_{1:d}^{1:N_x^{\textrm{PG}, 2}, -{\tilde{k}_{1:d}}}, y_{1:d})}.
	\end{align*}
	
	\subsection{Density Tempering}
	Again, recall that the $d$th target in a density tempering sequence of targets when using PMMH is
	\begin{align}
		\gamma_d(\params, x_{1:T}^{1:N_x^{\textrm{PMMH}}}, y_{1:T}) \ = \ p(\params)\widehat{p_{N_x^{\textrm{PMMH}}, d}}(y_{1:T}\mid\params, x_{1:T}^{1:N_x^{\textrm{PMMH}}}) \psi_d(x_{1:T}^{1:N_x^{\textrm{PMMH}}}\mid \params, y_{1:T}). \label{aeq:dt_pmmh_target_v2}
	\end{align} 
	and when using PG is 
	\begin{equation}
		\begin{aligned}
			&\gamma_d(\params, {k_{1:T}}, x_{1:T}^{k_{1:T}}, x_{1:T}^{1:N_x^{\textrm{PG}}, -{k_{1:T}}}, y_{1:T}) \\ &= \ p(\params)p(x_{1:T}^{k_{1:T}}\mid \params)\left[p(y_{1:T}\mid\params, x_{1:T}^{k_{1:T}}) \right]^{g_d} \psi_d(x_{1:T}^{1:N_x^{\textrm{PG}}, -{k_{1:T}}}\mid\params, y_{1:T}, x_{1:T}^{k_{1:T}}, {k_{1:T}}) \\ &\qquad \times \psi_d({k_{1:T}}\mid\params, y_{1:T}, x_{1:T}^{k_{1:T}}, x_{1:T}^{1:N_x^{\textrm{PG}}, -{k_{1:T}}}) \\
			&= \ p(\params)p(x_{1:T}^{k_{1:T}}\mid \params)\left[p(y_{1:T}\mid\params, x_{1:T}^{k_{1:T}}) \right]^{g_d} \psi_d(x_{1:T}^{1:N_x^{\textrm{PG}}, -{k_{1:T}}}, {k_{1:T}}\mid\params, y_{1:T}, x_{1:T}^{k_{1:T}}), \label{aeq:dt_pg_target}
		\end{aligned}
	\end{equation}
	Recall also that the density over the latent states is
	\begin{align*}
		p(x_{1:T}\mid\params) = \mu(x_1\mid\params)\prod_{t=2}^T{f(x_t\mid x_{t-1}, \params)},
	\end{align*}
	and the tempered density over the observations is
	\begin{align*}
		p_d(y_{1:T}\mid x_{1:T}, \params) = \prod_{t=1}^T{g(y_t\mid x_t, \params)^{g_d}} = \left(\prod_{t=1}^T{g(y_t\mid x_t, \params)}\right)^{g_d} = p(y_{1:T}\mid x_{1:T}, \params)^{g_d}.
	\end{align*}

	We denote
	\begin{align*}
		p_d(y_{1:T}\mid\params) = \int{p_d(x_{1:T}, y_{1:T}\mid\params)}dx_{1:T} = \int{p(x_{1:T}\mid\params)p_d(y_{1:T}\mid x_{1:T}, \params)}dx_{1:T}.
	\end{align*}
	The particle filter estimate $\widehat{p_{N_x^{\textrm{PMMH}}, d}}(y_{1:T}\mid\params, x_{1:T}^{1:N_x^{\textrm{PMMH}}})$, where $x_{1:T}^{1:N_x^{\textrm{PMMH}}}\sim\psi_d(\cdot\mid \params, y_{1:T})$ as described in Section \ref{sec:dt_targets}, is an unbiased estimate of $p_d(y_{1:T}\mid\params)$.

	\subsubsection{PG \texorpdfstring{$\rightarrow$}{->} PMMH}
	To reweight from \eqref{aeq:dt_pg_target} to \eqref{aeq:dt_pmmh_target_v2}, i.e.\ from PG to PMMH, the process is very similar to the data annealing case. First, a new set of latent state trajectories is drawn using the particle filter
	\begin{align*}
		x_{1:T}^{1:N_x^{\textrm{PMMH}}} \sim \psi_d(x_{1:T}^{1:N_x^{\textrm{PMMH}}}\mid \params, y_{1:T}),
	\end{align*}
	which gives the denominator of the kernel ratio in \eqref{eqn:kr} as 
	\begin{align*}
		&\int{\left[\gamma_d(\params, {k_{1:T}}, x_{1:T}^{k_{1:T}}, x_{1:T}^{1:N_x^{\textrm{PG}}, -{k_{1:T}}}, y_{1:T})\psi_d(x_{1:T}^{1:N_x^{\textrm{PMMH}}}\mid \params, y_{1:T})\right]}d\{{k_{1:T}}, x_{1:T}^{k_{1:T}}, x_{1:T}^{1:N_x^{\textrm{PG}}, -{k_{1:T}}}\} \\
		&= \int{\left[\gamma_d(\params, {k_{1:T}}, x_{1:T}^{k_{1:T}}, x_{1:T}^{1:N_x^{\textrm{PG}}, -{k_{1:T}}}, y_{1:T})\right]}d\{{k_{1:T}}, x_{1:T}^{k_{1:T}}, x_{1:T}^{1:N_x^{\textrm{PG}}, -{k_{1:T}}}\}\psi_d(x_{1:T}^{1:N_x^{\textrm{PMMH}}}\mid \params, y_{1:T}) \\
		&= p(\params)p_d(y_{1:T}\mid\params)\psi_d(x_{1:T}^{1:N_x^{\textrm{PMMH}}}\mid \params, y_{1:T}).
	\end{align*}
	The kernel ratio is 
	\begin{align*}
		\operatorname{KR}_d &= \frac{\gamma_d(\params, {k_{1:T}}, x_{1:T}^{k_{1:T}}, x_{1:T}^{1:N_x^{\textrm{PG}}, -{k_{1:T}}}, y_{1:T})}{p(\params)p_d(y_{1:T}\mid\params)\psi_d(x_{1:T}^{1:N_x^{\textrm{PMMH}}}\mid \params, y_{1:T})}.
	\end{align*}
	Replacing the exact likelihood $p_d(y_{1:T}\mid\params)$ with $\widehat{p_{N_x^{\textrm{PMMH}}, d}}(y_{1:T}\mid\params, x_{1:T}^{1:N_x^{\textrm{PMMH}}})$ gives 
	\begin{align*}
		\operatorname{KR}_d &= \frac{\gamma_d(\params, {k_{1:T}}, x_{1:T}^{k_{1:T}}, x_{1:T}^{1:N_x^{\textrm{PG}}, -{k_{1:T}}}, y_{1:T})}{\gamma_d(\params, x_{1:T}^{1:N_x^{\textrm{PMMH}}}, y_{1:T})}.
	\end{align*}
	
	\subsubsection{PMMH \texorpdfstring{$\rightarrow$}{->} PG}
	To reweight from \eqref{aeq:dt_pmmh_target_v2} to \eqref{aeq:dt_pg_target}, i.e.\ from PMMH to PG, we first run a particle filter with $N_x^{\textrm{PG}}$ state particles targeting $p(x_{1:T}\mid \params)\left[p(y_{1:T}\mid\params, x_{1:T}) \right]^{g_d}$, and then draw a new latent state trajectory,
	\begin{align*}
		({k_{1:T}}, x_{1:T}^{k_{1:T}}, x_{1:T}^{1:N_x^{\textrm{PG}}, -{k_{1:T}}})\sim\psi_d({k_{1:T}}, x_{1:T}^{k_{1:T}}, x_{1:T}^{1:N_x^{\textrm{PG}}, -{k_{1:T}}}\mid\params, y_{1:T}),
	\end{align*}
	where $\psi_d({k_{1:T}}, x_{1:T}^{k_{1:T}}, x_{1:T}^{1:N_x^{\textrm{PG}}, -{k_{1:T}}}\mid\params, y_{1:T})$ $=$ $\psi_d(x_{1:T}^{1:N_x^{\textrm{PG}}}\mid \params, y_{1:T})$ $\psi_d({k_{1:T}}\mid$$\params$$, $$y_{1:T}$$, $$x_{1:T}^{k_{1:T}}$$, $ $x_{1:T}^{1:N_x^{\textrm{PG}}, -{k_{1:T}}})$. 
	
	The kernel ratio in \eqref{eqn:kr} is then
	\begin{align*}
		\operatorname{KR}_d &= \frac{\gamma_d(\params, x_{1:T}^{1:N_x^{\textrm{PMMH}}}, y_{1:T})}{\int{\left[\gamma_d(\params, x_{1:T}^{1:N_x^{\textrm{PMMH}}}, y_{1:T})\psi_d({k_{1:T}}, x_{1:T}^{k_{1:T}}, x_{1:T}^{1:N_x^{\textrm{PG}}, -{k_{1:T}}}\mid\params, y_{1:T})\right]}d\{x_{1:T}^{1:N_x^{\textrm{PMMH}}}\}} \\
		&= \frac{\gamma_d(\params, x_{1:T}^{1:N_x^{\textrm{PMMH}}}, y_{1:T})}{p(\params)p_d(y_{1:T}\mid\params)\psi_d({k_{1:T}}, x_{1:T}^{k_{1:T}}, x_{1:T}^{1:N_x^{\textrm{PG}}, -{k_{1:T}}}\mid\params, y_{1:T})}.
	\end{align*}
	Again, by noting that 
	\begin{align*}
		&\psi_d(x_{1:T}^{1:N_x^{\textrm{PG}}, -{k_{1:T}}}, x_{1:T}^{k_{1:T}}, {k_{1:T}}\mid\params, y_{1:T}) \\
		&\qquad= \psi(x_{1:T}^{1:N_x^{\textrm{PG}}}\mid \params, y_{1:T})\psi_d({k_{1:T}}\mid\params, y_{1:T}, x_{1:T}^{k_{1:T}}, x_{1:T}^{1:N_x^{\textrm{PG}}, -{k_{1:T}}}) \\
		&= \psi_d(x_{1:T}^{1:N_x^{\textrm{PG}}, -{k_{1:d}}}, {k_{1:T}}\mid\params, y_{1:T}, x_{1:T}^{k_{1:T}})p_d(x_{1:T}^{k_{1:T}}\mid \params, y_{1:T}),
	\end{align*}
	the kernel ratio can be written as
	\begin{align*}
		\operatorname{KR}_d &= \frac{\gamma_d(\params, x_{1:T}^{1:N_x^{\textrm{PMMH}}}, y_{1:T})}{p(\params)p_d(y_{1:T}\mid\params)\psi_d(x_{1:T}^{1:N_x^{\textrm{PG}}, -{k_{1:d}}}, {k_{1:T}}\mid\params, y_{1:T}, x_{1:T}^{k_{1:T}})p_d(x_{1:T}^{k_{1:T}}\mid \params, y_{1:T})} \\
		&= \frac{\gamma_d(\params, x_{1:T}^{1:N_x^{\textrm{PMMH}}}, y_{1:T})}{\gamma_d(\params, {k_{1:T}}, x_{1:T}^{k_{1:T}}, x_{1:T}^{1:N_x^{\textrm{PG}}, -{k_{1:T}}}, y_{1:T})}.
	\end{align*}
	
	\subsubsection{PMMH \texorpdfstring{$\rightarrow$}{->} PMMH}
	To switch between two different PMMH kernels, i.e.\ to switch from $\gamma_d(\params, x_{1:T}^{1:N_x^{\textrm{PMMH}, 1}}, y_{1:T})$ to $\gamma_d(\params, x_{1:T}^{1:N_x^{\textrm{PMMH}, 2}}, y_{1:T})$, we draw a new set of latent state trajectories using a particle filter targeting $p(x_{1:T}\mid \params)\left[p(y_{1:T}\mid\params, x_{1:T}) \right]^{g_d}$
	\begin{align*}
		x_{1:T}^{1:N_x^{\textrm{PMMH}, 2}} \sim \psi_d(x_{1:T}^{1:N_x^{\textrm{PMMH}, 2}}\mid \params, y_{1:T}).
	\end{align*}
	Using this forward kernel, the kernel ratio in \eqref{eqn:kr} is
	\begin{align*}
		\operatorname{KR}_d &= \frac{\gamma_d(\params, x_{1:T}^{1:N_x^{\textrm{PMMH}, 1}}, y_{1:T})}{\int{\gamma_d(\params, x_{1:T}^{1:N_x^{\textrm{PMMH}, 1}}, y_{1:T})\psi_d(x_{1:T}^{1:N_x^{\textrm{PMMH}, 2}}\mid \params, y_{1:T})}dx_{1:T}^{1:N_x^{\textrm{PMMH}, 1}}} \\
		&= \frac{\gamma_d(\params, x_{1:T}^{1:N_x^{\textrm{PMMH}, 1}}, y_{1:T})}{p(\params)p_d(y_{1:T}\mid\params)\psi_d(x_{1:T}^{1:N_x^{\textrm{PMMH}, 2}}\mid \params, y_{1:T})}.
	\end{align*}
	As before, replacing the exact likelihood $p_d(y_{1:T}\mid\params)$ with the estimate $\widehat{p_{N_x^{\textrm{PMMH}, 2}, d}}(y_{1:T}\mid\params, x_{1:T}^{1:N_x^{\textrm{PMMH}, 2}})$ gives
	\begin{align*}
		\operatorname{KR}_d = \frac{\gamma_d(\params, x_{1:T}^{1:N_x^{\textrm{PMMH}, 1}}, y_{1:T})}{\gamma_d(\params, x_{1:T}^{1:N_x^{\textrm{PMMH}, 2}}, y_{1:T})}.
	\end{align*}

	See \citet{Botha2023} for a method to switch between two PMMH kernels targeting \eqref{eqn:dt_pmmh_target} (instead of \eqref{eqn:dt_pmmh_target_v2}).
	
	\subsubsection{PG \texorpdfstring{$\rightarrow$}{->} PG}
	To switch between two different PG kernels, i.e.\ to switch from $\gamma_d($$\params$$, $${k_{1:T}}$$, $$x_{1:T}^{k_{1:T}}$$, $$x_{1:T}^{1:N_x^{\textrm{PG}, 1}, -{k_{1:T}}},$ $y_{1:T})$ to $\gamma_d(\params, {\tilde{k}_{1:T}}, \tilde{x}_{1:T}^{\tilde{k}_{1:T}}, \tilde{x}_{1:T}^{1:N_x^{\textrm{PG}, 2}, -{\tilde{k}_{1:T}}}, y_{1:T})$, we run the conditional particle filter and
	draw a new latent state trajectory
	\begin{align*}
		({\tilde{k}_{1:T}}, \tilde{x}_{1:T}^{\tilde{k}_{1:T}}, \tilde{x}_{1:T}^{1:N_x^{\textrm{PG}, 2}, -{\tilde{k}_{1:T}}}) &\sim \psi_d(\tilde{x}_{1:T}^{1:N_x^{\textrm{PG}, 2}, -{\tilde{k}_{1:T}}}, {\tilde{k}_{1:T}}\mid\params, y_{1:T}, x_{1:T}^{k_{1:T}}).
	\end{align*}
	Then, the denominator of the kernel ratio in \eqref{eqn:kr} is
	\begin{align*}
		&\int{\gamma_d(\params, {k_{1:T}}, x_{1:T}^{k_{1:T}}, x_{1:T}^{1:N_x^{\textrm{PG}, 1}, -{k_{1:T}}}, y_{1:T})} \\
		&\qquad\times  {\psi_d(\tilde{x}_{1:T}^{1:N_x^{\textrm{PG}, 2}, -{\tilde{k}_{1:T}}}, {\tilde{k}_{1:T}}\mid\params, y_{1:T}, x_{1:T}^{k_{1:T}})}d\{{k_{1:T}}, x_{1:T}^{1:N_x^{\textrm{PG}, 1}, -{k_{1:T}}}\} \\
		&=\gamma_d(\params, {\tilde{k}_{1:T}}, \tilde{x}_{1:T}^{\tilde{k}_{1:T}}, \tilde{x}_{1:T}^{1:N_x^{\textrm{PG}, 2}, -{\tilde{k}_{1:T}}}, y_{1:T}).
	\end{align*}
	The kernel ratio is then 
	\begin{align*}
		\operatorname{KR}_d = \frac{\gamma_d(\params, {k_{1:T}}, x_{1:T}^{k_{1:T}}, x_{1:T}^{1:N_x^{\textrm{PG}, 1}, -{k_{1:T}}}, y_{1:T})}{\gamma_d(\params, {\tilde{k}_{1:T}}, \tilde{x}_{1:T}^{\tilde{k}_{1:T}}, \tilde{x}_{1:T}^{1:N_x^{\textrm{PG}, 2}, -{\tilde{k}_{1:T}}}, y_{1:T})}.
	\end{align*}
	
\end{document}